\documentclass[12pt, a4, reqno]{amsart}
\usepackage{natbib}
\usepackage{amsmath,amssymb,centernot,mathtools}
\usepackage{amsthm}
\usepackage{graphicx,subcaption}
\usepackage{caption}
\usepackage{enumitem}
\usepackage{color}
\usepackage[table]{xcolor}
\usepackage[calc]{picture}
\usepackage[english]{babel}
\usepackage[margin=1in]{geometry}
\usepackage{secdot}
\usepackage{float}
\usepackage{tikz}
\usetikzlibrary{arrows.meta,shapes}
\usetikzlibrary{arrows,shapes.arrows,shapes.geometric,shapes.multipart,
decorations.pathmorphing,positioning,shapes.swigs,}

\usepackage{setspace}
\usepackage{titlesec}
\usepackage{tcolorbox}
\usepackage[outline]{contour}
\usepackage[markers,nolists]{endfloat}
\usepackage{amsfonts}
\usepackage{algcompatible}
\usepackage{algorithm,algpseudocode,float}
\usepackage{pdflscape}
\usepackage{cancel}
\usepackage{xr}
\usepackage{etoolbox}

\renewenvironment{titlepage}{%
  \setcounter{page}{0}
  \thispagestyle{empty}
  \centering
  \vspace*{\fill}
}{%
  \vspace{0\baselineskip}
  \vspace*{\fill}
  \newpage
}
\makeatletter
\patchcmd{\@maketitle}
{\if@titlepage \newpage \else}
{\if@titlepage
 \vspace{\baselineskip}
 \else}
{}{}
\makeatother

\newtheorem{theorem}{Theorem}
\newtheorem{assumption}{Assumption}

\newtheorem{example}{Example} 

\DeclareMathOperator*{\expit}{expit}
\DeclareMathOperator*{\eff}{eff}

\newcommand{\CI}{\mathrel{\perp\mspace{-10mu}\perp}}
\newcommand{\nCI}{\centernot{\CI}}

\makeatletter
\g@addto@macro\normalsize{%
  \setlength\abovedisplayskip{10pt}
  \setlength\belowdisplayskip{10pt}
  \setlength\abovedisplayshortskip{10pt}
  \setlength\belowdisplayshortskip{10pt}
}
\makeatother

\titlespacing*{\section}
  {0pt}{1\baselineskip}{1\baselineskip}

\makeatletter
\newenvironment{breakablealgorithm}
  {
   \begin{center}
     \refstepcounter{algorithm}
     \hrule height.8pt depth0pt \kern2pt
     \renewcommand{\caption}[2][\relax]{
       {\raggedright\textbf{\ALG@name~\thealgorithm} ##2\par}%
       \ifx\relax##1\relax 
         \addcontentsline{loa}{algorithm}{\protect\numberline{\thealgorithm}##2}%
       \else 
         \addcontentsline{loa}{algorithm}{\protect\numberline{\thealgorithm}##1}%
       \fi
       \kern2pt\hrule\kern2pt
     }
  }{
     \kern2pt\hrule\relax
   \end{center}
  }
\makeatother

\doublespace

\begin{document}


\begin{titlepage}
\pagenumbering{gobble}

\title[Intervening variables and unmeasured confounding]{Causal effects of intervening variables in settings with unmeasured confounding}

\author[Lan Wen, Aaron L. Sarvet and Mats J. Stensrud]{Lan Wen$^{\text{a}}$, Aaron L. Sarvet$^{\text{b}}$ \and Mats J. Stensrud$^{\text{b}}$ \vspace{0.6cm}
\\ \small
$^\text{a}$Department of Statistics and Actuarial Science, University of Waterloo, Waterloo, Ontario, Canada \vspace{0.3cm} \\
$^\text{b}$Department of Mathematics, Ecole Polytechnique F\'{e}d\'{e}rale de Lausanne, Lausanne, Switzerland\\}

\maketitle

\begin{abstract}
We present new results on average causal effects in settings with unmeasured exposure-outcome confounding. Our results are motivated by a class of estimands, e.g.,\ frequently of interest in medicine and public health, that are currently not targeted by standard approaches for average causal effects. We recognize these estimands as queries about the average causal effect of an \textit{intervening} variable. We anchor our introduction of these estimands in an investigation of the role of chronic pain and opioid prescription patterns in the opioid epidemic, and illustrate how conventional approaches will lead unreplicable estimates with ambiguous policy implications. We argue that our altenative effects are replicable and have clear policy implications, and furthermore are non-parametrically identified by the classical frontdoor formula. As an independent contribution, we derive a new semiparametric efficient estimator of the frontdoor formula with a uniform sample boundedness guarantee. This property is unique among previously-described estimators in its class, and we demonstrate superior performance in finite-sample settings. Theoretical results are applied with data from the National Health and Nutrition Examination Survey.

\end{abstract}

\end{titlepage}

\pagenumbering{arabic}

\section{Introduction}
\label{sec:intro}

Unmeasured confounding and ill-defined interventions are major contributing factors to the replication crisis for policy-relevant parameters, e.g.,\ in medical research. When there is unmeasured confounding, standard covariate-adjustment approaches will lead to biases that would likely differ across studies. Further, when an analysis is based on variables that do not correspond to well-defined interventions, it is nearly impossible for future analyses and experiments to ensure that these variables are operationalized identically with the original study. In each case, replication is nearly impossible.  

To counteract these challenges, there exists a diverse set of strategies to confront unmeasured confounding between exposure and outcome, which leverage the measurement of auxiliary variables, including instruments and other proxies \citep{angrist1996identification, lipsitch2010negative,  tchetgen2020introduction}, as well as mediators \citep{pearl2009causality,Fulcher2020}. At the same time, there is a long history of calls for an ``interventionist'' approach to causal analyses in statistics \citep{holland1986statistics, dawid2000causal, richardson2013single, robins2022interventionist} wherein investigators focus on the effects of manipulable, or \textit{intervenable}, variables.  Use of such variables ensures that the targets of inference are clearly defined by variables representing interventions that can be implemented in principle (see e.g., \citealp{hernan2005invited, hernan2011compound, galea2013argument}). Furthermore, the seemingly disparate challenges of ill-defined interventions and unmeasured exposure-outcome confounding are often considered separately, but systematically co-occur in practice: when an exposure variable does not correspond to a well-defined intervention, investigators will often face exceptional challenges in sufficiently describing and measuring the causes of that exposure. In such cases, investigators will often also have little confidence in the assumption of no unmeasured confounding \citep{ hernan2008does}.  

In this article we consider jointly the twin challenges of ill-defined interventions and unmeasured confounding. Building on new results for a generalized theory of separable effects \citep{robins2022interventionist,Robins2010,stensrud2021generalized}, our contributions concern effects of an \textit{intervening} variable: a manipulable descendant of a (possibly ill-defined) exposure (or treatment) that precedes the outcome. We argue that such average causal effects, rather than those of (possibly ill-defined) exposures, are frequently of interest in practice. We give results on their interpretation, identification and estimation. In doing so, we develop a novel semiparametric efficient estimator for the canonical front-door functional with superior finite sample performance properties.  

\subsection{Related approaches}
\label{sec:related work}
Our results are related to previous work on identification of causal effects in the presence of unmeasured confounding. As we expound in Section \ref{sec:manip_simple}, causal effect of an \textit{intervening} variable may be identified by the frontdoor formula, which coincidentally also allows non-parametric identification of the average causal effect (ACE) of exposure on outcome, even in the presence of unmeasured exposure-outcome confounding \citep {pearl1993mediating, pearl2009causality}. Yet meaningful applications of the frontdoor criterion to study the ACE of exposure on outcome have been scarce. One problem is that conventional identification of the ACE by the frontdoor criterion requires a strict exclusion restriction, which is often infeasible: an investigator must measure at least one mediator intersecting each causal pathway from the exposure to the outcome. In contrast to conventional approaches, however, the definition of \textit{intervening} variables often renders such exclusion restrictions uniquely plausible.

Our results are also related to the work by \citet{Fulcher2020}, who gave conditions under which the frontdoor formula identifies the so-called Population Intervention Indirect Effect (PIIE). \citet{Fulcher2020} interpreted the PIIE as a ``contrast between the observed outcome mean for the population and the population outcome mean if contrary to fact the mediator had taken the value that it would have in the absence of exposure.'' Thus, unlike the conventional identification result for the ACE, the frontdoor formula can be used to identify the PIIE even in the presence of a direct effect of the exposure on the outcome not mediated by an intermediate variable (or intermediate variables). We describe the assumptions for the two aforementioned estimands in Section \ref{sec:idassumptions}. To fix ideas about the relation to the previous work on the frontdoor formula, we introduce the following running example. 

\begin{example}[Chronic pain and opioid use, \citealp{inoue2022causal}]
\label{ex:inoue}
\normalfont 
Chronic pain is associated with use and overdosing of opioids, which subsequently can lead to death. Moreover, chronic pain can also affect mortality outside of its effect on opioid use \citep{dowell2016cdc}, e.g.,\ by causing long-term stress and undesirable lifestyle changes. \citet{inoue2022causal} studied the effect of chronic pain  (exposure to pain versus no pain) on mortality (outcome) mediated by opioid use. Chronic pain is notoriously treatment-resistant, and unmeasured confounders between chronic pain and mortality may include social, physiologic, and psychological factors \citep{inoue2022causal}.
Using data from the National Health and Nutrition Examination Survey (NHANES) from 1999--2004 with linkage to mortality databases through 2015, investigators studied a causal effect related to the PIIE. Specifically, \citet{inoue2022causal} considered a ``path-specific frontdoor effect" that they interpreted as ``the change in potential outcomes that follows a change in the mediator (opioid) which was caused by changing the exposure." 
\end{example}

As indicated by \citet{inoue2022causal} there likely exist unmeasured common causes of the exposure and the outcome in our example. 
However, it is unclear how one could intervene on the exposure variable, chronic pain, or even whether any intervention on chronic pain can possibly be well-defined. Therefore, the estimands in our example have dubious public health implications \citep{holland1986statistics}. In contrast, we will suggest effects of \textit{intervening} variables, which do correspond to interventions that are feasible to implement in practice. To motivate our intervention, consider a doctor who determines if a patient should receive opioids to relieve symptoms. The doctor's decision could be affected by whether the patient has chronic pain. However, the current guidelines by the Centers for Disease Control and Prevention (CDC) \citep{dowell2016cdc,inoue2022causal} suggests that chronic pain status should no longer be used to determine opioid assignment. Thus, there is interest in evaluating outcomes under a modified prescription policy (the intervention of interest), such that the doctor disregards a patient's chronic pain in their decisions on opioid prescriptions. To implement this modified prescription policy, doctors could e.g.,\ be asked to not consider a patient as having chronic pain in the prescription process. This modified policy does not require us to conceptualize interventions on a patient's chronic pain -- an intervention that would have been hard, or impossible, to specify. However, an analyst with access to observational data might reasonably assume a deterministic relation between a patient's chronic pain status and the \textit{doctor's perception} of the chronic pain status: an \textit{intervening} variable that in turn determines the doctor's prescription in the observed data. 


The remainder of the articles is organized as follows. In Section \ref{sec:obs data} we describe the observed and counterfactual data structure. In Section \ref{sec:manip_simple} we precisely define our interventionist estimand and derive identification results. In Section \ref{sec:idassumptions} we relate our identification results to non-parametric conditions that allow identification of relevant estimands that have been proposed in the past, and discuss the plausibility of these conditions.  In Section \ref{sec:estimators} we present a new semiparametric estimator of the frontdoor formula. 
In Section \ref{sec:sims} we give simulation results and illustrate that our estimator performs well in finite sample settings. In Section \ref{sec:application}, we apply our new results to study the effect of opioid prescription policies on chronic pain using data from NHANES. We discuss the practical implication of our work in Section \ref{sec:discussion}.

\section{Observed and counterfactual data structure}
\label{sec:obs data}
Consider a study of $n$ iid individuals randomly sampled from a large superpopulation. Let $A$ denote the observed binary exposure taking values $a^\dagger$ or $a^\circ$ (e.g., chronic pain). $M$ (e.g., opioid usage),  a mediator variable and $Y$, an outcome of interest which can be binary, categorical or continuous (e.g., probability of survival at $3$ years or five years).  Furthermore, suppose that $L$ is a vector of pre-exposure covariates measured at baseline, which can include common causes of $A,~M$ and $Y$. To simplify the presentation, we will assume throughout that all covariates are discrete, that is, the variables have distributions that are absolutely continuous with respect to a counting measure. However, our arguments extend to settings with continuous covariates and the Lebesgue measure. We indicate counterfactuals in superscripts.  In particular, let $M^{a}$ and $Y^{a}$ denote the counterfactual mediator and outcome variables if, possibly contrary to fact, the exposure had taken a value $a$ for $a\in \{a^\dagger, a^\circ\}$. 
Extensions to discrete exposure variables with more than two levels are discussed in Appendix \ref{sec:appendix_discrete}.


\section{Effects of the intervening variable}
\label{sec:manip_simple}


In the analysis of the chronic pain example,  \citet{inoue2022causal} concluded that their results are relevant to pain management, arguing that the findings ``highlight the importance of careful guideline-based chronic pain management to prevent death from possibly inappropriate opioid prescriptions driven by chronic pain." However, this argument does not translate to an intervention on chronic pain $A$; rather, we interpret their policy concern as one that directly involves a modifiable \textit{intervening} variable: a care provider's \textit{perception} of the patient's chronic pain in standard-of-care pain management decisions, say $A_M$. If we define $A_M$ as binary (taking values $a^\dagger$ or $a^\circ$), and assume that a care provider's natural consideration corresponds exactly with a patient's chronic pain experience, then in this setting, $A = A_M$ with probability one in the observed data. Despite this feature of the observed data, we could nevertheless conceive an intervention that modifies this \textit{intervening} variable $A_M$ without changing the non-modifiable exposure $A$, and thus preserve the investigators direct policy concerns at this causal estimand-formulation stage of the analysis. 
Our consideration of the chronic pain example motivates estimands under interventions on the modifiable \textit{intervening} variable $A_M$, not $A$, where (i) $A_M$ is deterministically equal to $A$ in the observed data, and (ii) $A_M$ captures the effects of $A$ on $Y$ through $M$. These two features of our estimand are also features of separable effects \citep{Robins2010, robins2020interventionist,stensrud2021generalized}.

{\bf Definition 1. }(The intervening variable estimand)
The \textit{intervening} variable estimand is defined as the expected counterfactual outcome $E(Y^{a_M})$, where the intervening variable $a_M$ can take values $a^\dagger$ or $a^\circ$ in the sample space of $A$.

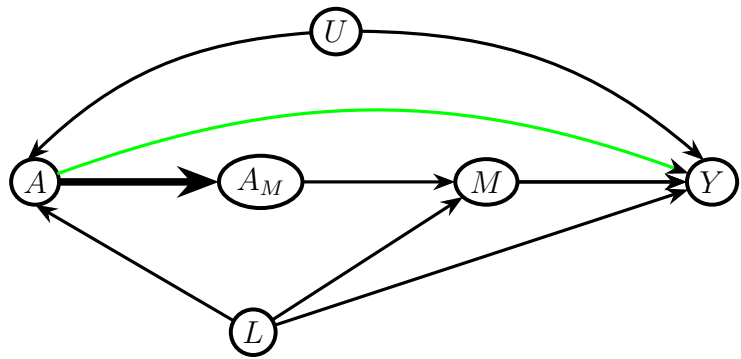
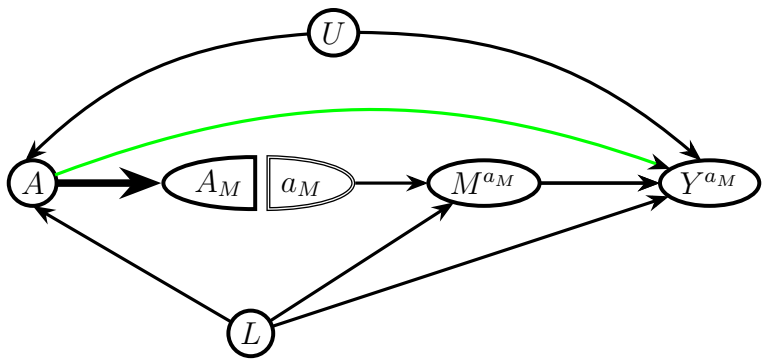
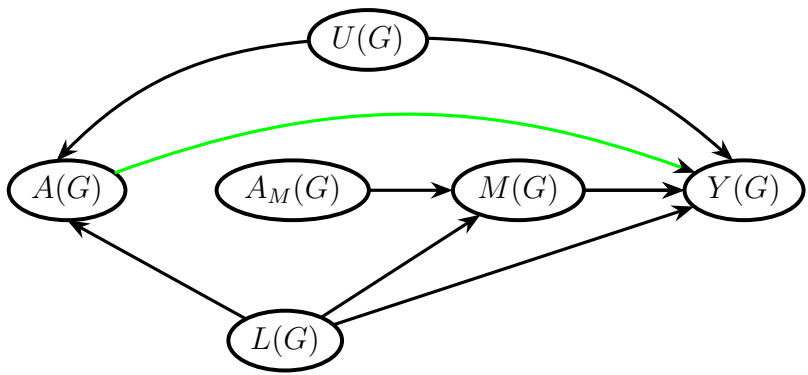
\begin{figure}
    \centering
\begin{minipage}{1\textwidth}
      \centering
                \begin{tikzpicture}
                    \tikzset{line width=1.5pt, outer sep=0pt,
                    ell/.style={draw,fill=white, inner sep=2pt,
                    line width=1.5pt},
                    swig vsplit={gap=5pt,
                    inner line width right=0.5pt}};
                    \node[name=AM,ell,  shape=ellipse] at (-3,0) {$A_M$};
                    \node[name=A,ell,  shape=ellipse] at (-6,0) {$A$};
                    \node[name=M,ell,  shape=ellipse] at (0,0) {$M$};
                    \node[name=H,ell,  shape=ellipse] at (-2,2) {$U$};
                    \node[name=Y,ell,  shape=ellipse] at (3,0) {$Y$};
                    \node[name=L,ell,  shape=ellipse] at (-3.1, -2) {$L$};

                    \begin{scope}[>={Stealth[black]},
                                  every edge/.style={draw=black,very thick}]
                        \path [->] (A) edge[line width=1mm] (AM);
                        \path [->] (M) edge (Y);
                        \path [->] (AM) edge (M);
                        \path [->] (H) edge[bend right=20] (A.110);
                        \path [->] (H) edge[bend left=20] (Y.110);
                        \path [->] (M) edge (Y);
                        \path [->] (A) edge[bend right=-20, color=green] (Y);      
                        \path [->] (L) edge (A.270);
                        \path [->] (L) edge (M);
                        \path [->] (L) edge (Y);

                    \end{scope}
                \end{tikzpicture}
\subcaption{\label{fig:front6a}Modified extended DAG.}
\end{minipage} \\ \vspace{2em}
\begin{minipage}{1\textwidth}
      \centering
                \begin{tikzpicture}
                    \tikzset{line width=1.5pt, outer sep=0pt,
                    ell/.style={draw,fill=white, inner sep=2pt,
                    line width=1.5pt},
                    swig vsplit={gap=5pt,
                    inner line width right=0.5pt}};
                    \node[name=AM,shape=swig vsplit] at (-3,0){
                    \nodepart{left}{$A_M$}
                    \nodepart{right}{$a_M$} };
                    \node[name=A,ell,  shape=ellipse] at (-6,0) {$A$};
                    \node[name=M,ell,  shape=ellipse] at (0,0) {$M^{a_M}$};
                    \node[name=H,ell,  shape=ellipse] at (-2,2) {$U$};
                    \node[name=Y,ell,  shape=ellipse] at (3,0) {$Y^{a_M}$};
                     \node[name=L,ell,  shape=ellipse] at (-3.1, -2) {$L$};
                    \begin{scope}[>={Stealth[black]},
                                  every edge/.style={draw=black,very thick}]
                        \path [->] (A) edge[line width=1mm] (AM);
                        \path [->] (M) edge (Y);
                        \path [->] (AM) edge (M);
                        \path [->] (H) edge[bend right=20] (A.110);
                        \path [->] (H) edge[bend left=20] (Y.110);
                        \path [->] (M) edge (Y);
                        \path [->] (A) edge[bend right=-20, color=green] (Y);   
                        \path [->] (L) edge (A.270);
                        \path [->] (L) edge (M);
                        \path [->] (L) edge (Y);

                    \end{scope}
                \end{tikzpicture}
\subcaption{\label{fig:front6b} SWIG corresponding to an intervention on the modified extended graph on in Figure \ref{fig:front6a}.}
\end{minipage}\\ \vspace{2em}
\begin{minipage}{1\textwidth}
      \centering
                \begin{tikzpicture}
                    \tikzset{line width=1.5pt, outer sep=0pt,
                    ell/.style={draw,fill=white, inner sep=2pt,
                    line width=1.5pt},
                    swig vsplit={gap=5pt,
                    inner line width right=0.5pt}};
                \node[name=AM,ell,  shape=ellipse] at (-3,0){$A_M(G)$};
                    \node[name=A,ell,  shape=ellipse] at (-6,0) {$A(G)$};
                    \node[name=M,ell,  shape=ellipse] at (0,0) {$M(G)$};
                    \node[name=H,ell,  shape=ellipse] at (-2,2) {$U(G)$};
                    \node[name=Y,ell,  shape=ellipse] at (3,0) {$Y(G)$};
                    \node[name=L,ell,  shape=ellipse] at (-3.1, -2) {$L(G)$};
                    \begin{scope}[>={Stealth[black]},
                                  every edge/.style={draw=black,very thick}]
                        \path [->] (M) edge (Y);
                        \path [->] (AM) edge (M);
                        \path [->] (H) edge[bend right=20] (A.110);
                        \path [->] (H) edge[bend left=20] (Y.110);
                        \path [->] (M) edge (Y);
                        \path [->] (A) edge[bend right=-20, color=green] (Y); 
                        \path [->] (L) edge (A.270);
                        \path [->] (L) edge (M);
                        \path [->] (L) edge (Y);


                    \end{scope}
                \end{tikzpicture}
\subcaption{\label{fig:front6c} DAG for reading off Assumption \ref{ass:dismiss}.}
\end{minipage}
\caption{DAG and SWIGs of modified extended graph.}
\end{figure}

As such, we can define a causal effect of an \textit{intervening} variable $A_M$ on an outcome $Y$ as a contrast between $E(Y^{a_M=a^\dagger})$ and another causal estimand such as $E(Y^{a_M=a^\circ})$ or $E(Y)$.
As in the separable effects literature, the conceptual elaboration in the chronic pain example is amenable to graphical representation. Consider an extended causal directed acyclic graph (DAG) \citep{Robins2010}, which not only includes $A$ but also the \textit{intervening} variable $A_M$.  
Figure \ref{fig:front6a} shows this extended DAG with node set $V = (U,L,A,A_M,M,Y)$. The bold arrow from $A$ to $A_M$ in Figure \ref{fig:front6a} indicates a deterministic relationship in the observed data. The deterministic relationship encodes that, in the observed data, with probability one under $f(v)$, either  $A = A_M = a^\dagger$ or $A = A_M  = a^\circ$. These graphs will be used to illustrate our subsequent results, where we give formal conditions under which the effects of an intervention that sets $A_M$ to $a_M$, as represented in the Single World Interventions Graph (SWIG) Figure \ref{fig:front6b}, can be identified.
Note that the absence of an arrow in a SWIG encodes the absence of individual level effects as described in \citet{richardson2013single}.
Causal effects of {intervening} variables are of interest beyond our running chronic pain example. We illustrate another application in the following example considering race (an exposure often under-theorized in statistical analyses) and job interview discrimination. In Appendix \ref{eg:fulcher}, we also consider an obstetrics example from \citet{Fulcher2020}.

\begin{example}[Race and job interview discrimination]
\normalfont 
Consider the extended graph in Figure \ref{fig:front6a}, where $A$ denotes the race of an individual, $A_M$ denotes the race that an individual indicates on a job application for a company, $M$ denotes the indicator that the individual receives an interview, and $Y$ denotes whether an individual is hired for the job (1 if the individual is hired and 0 otherwise). Further, the unmeasured variable $U$ includes complex historical processes that ultimately affect both a person's perceived or declared race on the job market and may influence whether or not they are hired by the company. Several studies have documented that resum\'e `whitening' (e.g., deleting any references or connotations to a non-White race) can increase an applicant's chance of receiving an interviews \citep{Kang2016, Gerdeman2017}. However, it is unclear if this strategy leads to a higher chance of actually being hired, as e.g., unconscious bias can also affect whether or not an individual is hired, even if the candidate's background and qualifications are the same as other candidates. Thus, in a group of individuals with the same qualifications, the difference between $E(Y)$ and $E(Y^{a_M=\text{white}})$ could indicate the effect among non-White candidates of resum\'e whitening in the screening process on the probability of being hired.
\end{example}

To formally state our identifiability conditions of an intervening variable estimand, we first invoke the assumption of a deterministic relationship between the exposure and the intervening variable in the observed data. 
\begin{assumption}[Intervening variable determinism]
\label{ass:det}
$A = A_M$ \text{w.p.1.}
\end{assumption}

We also invoke a positivity condition which only involves observable laws and requires that for all joint values of $L$, there is a positive probability of observing $A=a$ and $M=m$, $\forall a,m$. 

\begin{assumption}[Positivity]
\label{ass:pos}
   $f(m, a \mid l)>0$, $\forall (m,~a,~l)\in\textup{supp}({M},~{A},~{L})$. 
\end{assumption}

Further, we use a consistency assumption stating that the interventions on the {intervening} variable  $A_M$ is well-defined.
\begin{assumption}[Consistency] If $A_M=a_M$, then
\label{ass:consG}
$M^{a_M} = M,
Y^{a_M} = Y,  ~~\forall ~a_M\in\textup{supp}({A}).$
\end{assumption}

Following \citet{stensrud2021generalized}, let ``$(G)$"  refer to a future trial where $A_M$ is randomly assigned, and consider the following dismissible component conditions. 
\begin{assumption}[Dismissible component conditions]
\label{ass:dismiss}
\begin{align}
& Y(G) \CI A_M(G) \mid A(G),  L(G), M(G) \label{ass: delta 1}, \\
& M(G) \CI A(G) \mid A_M(G),  L(G) \label{ass: delta 2}.
\end{align} 
 \end{assumption}
  Assumption \ref{ass:dismiss} can fail either when conditional independence \eqref{ass: delta 1} fails (e.g., when $A_M$ exerts effects on $Y$ not intersected by $M$) or when conditional independence \eqref{ass: delta 2} fails (e.g., when $A$ exerts effects on $M$ not intersected by $A_M$). Furthermore, Assumption \ref{ass:dismiss} will  require that $L$ includes certain common causes of other variables; it is sufficient that $L$ captures all common causes of $M$ and $Y$ and all common causes of $A$ and $M$.
Note that the dismissible component conditions imply so-called partial isolation conditions (see \citealp{stensrud2021generalized}).

\begin{theorem}
    The average counterfactual outcome under an intervention on $A_M$ is identified by the frontdoor formula \eqref{eq:gen_frontdoor} under Assumptions \ref{ass:det}-\ref{ass:dismiss}. That is,
    \begin{align}
\Psi &\coloneqq E(Y^{a_M=a^\dagger})\label{eq:gen_frontdoor}
\\& = \sum_{m,l} f(m\mid a^\dagger,l) f(l)\sum_a E(Y\mid L=l,A=a, M=m)f(a\mid l).
\nonumber
\end{align}
\end{theorem}
As we formally show in Appendix \ref{app:sepeff_generalid}, our causally manipulable estimand $E(Y^{a_M=a^\dagger})$ is identified by \eqref{eq:gen_frontdoor} even when $A$ is a direct cause of $Y$, not mediated  by $M$. Henceforth, we refer to the right-hand-side of \eqref{eq:gen_frontdoor} as the generalized frontdoor formula as it allows for some baseline covariates $L$. To the best of our knowledge, Pearl's original frontdoor formula \citep{pearl1993mediating, pearl2009causality} did not consider $L$. Note that \citet{Fulcher2020} used the term generalized frontdoor criterion (not formula) to describe the identification conditions for the PIIE, which we discuss in Section \ref{sec:idassumptions}. They argued that their ``identification criterion generalizes Judea Pearl's front door criterion as it does not require no direct effect of exposure not mediated by the intermediate variable". However, the cross-world exchangeability Assumption \ref{ass:exchCross}  is necessary to identify PIIE, but it is not necessary to identify the ACE. Thus, \citet{Fulcher2020} added  an (untestable) assumption that is unnecessarily restrictive when the aim is to identify the ACE.

To motivate our estimation results in Section \ref{sec:estimators}, we emphasize that, in the absence of covariates $L$, the frontdoor formula can be expressed as
\begin{align}
\Psi &\coloneqq E(Y^{a_M = a^\dagger})  = \sum_{m} f(m\mid a^\dagger) \sum_a E(Y\mid A=a, M=m)f(a)
\label{eq: simple frontdoor formula}\\ & = P(A=a^\dagger) E(Y\mid A=a^\dagger) 
 + P(A=a^\circ) \sum_{m} E(Y\mid A=a^\circ, M=m) f(m\mid a^\dagger).
 \nonumber
\end{align} 
Thus, $E(Y^{a_M = a^\dagger})$ is a weighted average of a conditional mean, $E(Y\mid A=a^\dagger)$, and a sum that appears in known identification formulae for both natural (pure) and separable effects of exposure or treatment on $Y$ \citep{Robins2010,robins2020interventionist,stensrud2021generalized,stensrud2022conditional}. The decomposition of the frontdoor formula in \eqref{eq: simple frontdoor formula} is instrumental in formulating new semiparametric estimators (see Appendix \ref{sec: app proof frontdoor} for further details). In Appendix \ref{sec:appendix_nongen}, we give identification and estimation results for our new causally manipulable estimand in absence of $L$, and in Appendix \ref{app:sepeff_generalid}, we provide further intuition for the  identification results based on a weighted decomposition of the frontdoor formula.

\section{Other estimands that can be identified using the frontdoor formula}
\label{sec:idassumptions}
To give context to our identification results, we review sufficient assumptions ensuring that the frontdoor formula identifies the ACE and the PIIE \citep{pearl2009causality,Fulcher2020}. For exposure levels $a^\dagger$ and $a^\circ$, the ACE and PIIE are given respectively by
\begin{align*}
    E(Y^{a^\dagger}) \ \text{vs.}\   E(Y^{a^\circ}),~~\text{and}~~
    E(Y) \ \text{vs.} \ E(Y^{M^{a^\dagger},A}).
\end{align*}

Beyond the Assumption \ref{ass:pos} (Positivity), consider the following assumptions.
\begin{assumption}[Consistency] If $A=a$ and $M=m$, then
\label{ass:cons2}
$$
M^{a} = M, \quad
Y^{a} = Y, \quad
Y^{a,m} = Y, \ \forall (m,~a)\in\textup{supp}({M},~{A}).
$$
\end{assumption}
\begin{assumption}[Exposure - Mediator Exchangeability]
   \label{ass:exchAM}
   $M^a \CI A \mid L$,  $\forall ~a ~ \in ~ \textup{supp}({A})$ .
\end{assumption}
Consistency Assumption \ref{ass:cons2} requires well-defined interventions on both $A$ and $M$, which seems to be implausible in our example, and Exchangeability Assumption \ref{ass:exchAM} ensures that the exposure-mediator association is unconfounded, which holds for instance in the SWIG in Figure \ref{fig:front2b}. Both assumptions are used in the identification of the ACE and the PIIE in \citet{Fulcher2020}, but are not strictly necessary for frontdoor identification of the ACE (see \citealp{didelez2018causal} for assumptions without intervention on $M$).

\subsection{Identification of the ACE}

In addition to Assumptions \ref{ass:pos}, \ref{ass:cons2} and \ref{ass:exchAM}, the following two conditions are sufficient for the identification of the ACE using the generalized frontdoor formula \eqref{eq:gen_frontdoor}.

\begin{assumption}[No Direct Effect] 
\label{ass:noDirect}
$Y^{a,m} = Y^m.$
\end{assumption}

\begin{figure}
    \begin{minipage}{1\textwidth}
        \centering
                \begin{tikzpicture}
                    \tikzset{line width=1.5pt, outer sep=0pt,
                    ell/.style={draw,fill=white, inner sep=2pt,
                    line width=1.5pt},
                    swig vsplit={gap=5pt,
                    inner line width right=0.5pt}};
                    \node[name=A,ell,  shape=ellipse] at (-3,0) {$A$};
                    \node[name=M,ell,  shape=ellipse] at (0,0) {$M$};
                    \node[name=H,ell,  shape=ellipse] at (0,2) {$U$};
                    \node[name=Y,ell,  shape=ellipse] at (3,0) {$Y$};
                    \node[name=L,ell,  shape=ellipse] at (-3, -2) {$L$};

                    \begin{scope}[>={Stealth[black]},
                                  every edge/.style={draw=black,very thick}]
                        \path [->] (M) edge (Y);
                        \path [->] (A) edge (M);
                        \path [->] (H) edge[bend right] (A.110);
                        \path [->] (H) edge[bend left] (Y.110);
                        \path [->] (M) edge (Y);
                        \path [->] (A) edge[bend right=-30, color=green] (Y);    
                        \path [->] (L) edge (A);
                        \path [->] (L) edge (M);
                        \path [->] (L) edge (Y);
                    \end{scope}
                \end{tikzpicture}
\subcaption{\label{fig:front1}DAG of the observed random variables.}
\end{minipage}
\\  \vspace{2em}
\begin{minipage}{1\textwidth}
    \centering
                \begin{tikzpicture}
                    \tikzset{line width=1.5pt, outer sep=0pt,
                    ell/.style={draw,fill=white, inner sep=2pt,
                    line width=1.5pt},
                    swig vsplit={gap=5pt,
                    inner line width right=0.5pt}};
                    \node[name=A,shape=swig vsplit] at (-3,0){
                    \nodepart{left}{$A$}
                    \nodepart{right}{$a$} };
                    \node[name=M,ell,  shape=ellipse] at (0,0) {$M^a$};
                    \node[name=H,ell,  shape=ellipse] at (0,2) {$U$};
                    \node[name=Y,ell,  shape=ellipse] at (3,0) {$Y^a$};
                     \node[name=L,ell,  shape=ellipse] at (-3.1, -2) {$L$};
                    \begin{scope}[>={Stealth[black]},
                                  every edge/.style={draw=black,very thick}]
                        \path [->] (M) edge (Y);
                        \path [->] (A) edge (M);
                        \path [->] (H) edge[bend right] (A.110);
                        \path [->] (H) edge[bend left] (Y.110);
                        \path [->] (M) edge (Y);
                        \path [->] (A) edge[bend right=-30, color=green] (Y);      
                        \path [->] (L) edge (A.270);
                        \path [->] (L) edge (M);
                        \path [->] (L) edge (Y);

                    \end{scope}
                \end{tikzpicture}
\subcaption{\label{fig:front2a}SWIG of the counterfactual variables corresponding to Figure \ref{fig:front1} with intervention on treatment or exposure $A$.}
\end{minipage}\\  \vspace{2em}
\begin{minipage}{1\textwidth}
\centering
                \begin{tikzpicture}
                    \tikzset{line width=1.5pt, outer sep=0pt,
                    ell/.style={draw,fill=white, inner sep=2pt,
                    line width=1.5pt},
                    swig vsplit={gap=5pt,
                    inner line width right=0.5pt}};
                    \node[name=A,ell,  shape=ellipse] at (-3,0) {$A$};
                    \node[name=M,shape=swig vsplit] at (0,0){
                    \nodepart{left}{$M$}
                    \nodepart{right}{$m$} };
                    \node[name=H,ell,  shape=ellipse] at (0,2) {$U$};
                    \node[name=Y,ell,  shape=ellipse] at (3,0) {$Y^m$};
                    \node[name=L,ell,  shape=ellipse] at (-3.1, -2) {$L$};
                    \begin{scope}[>={Stealth[black]},
                                  every edge/.style={draw=black,very thick}]
                        \path [->] (M) edge (Y);
                        \path [->] (A) edge (M);
                        \path [->] (H) edge[bend right] (A);
                        \path [->] (H) edge[bend left] (Y);
                        \path [->] (M) edge (Y);
                        \path [->] (A) edge[bend right=-30, color=green] (Y);    

                        \path [->] (L) edge (A.270);
                        \path [->] (L) edge (M);
                        \path [->] (L) edge (Y);

                    \end{scope}
                \end{tikzpicture} 
\subcaption{\label{fig:front2b}SWIG of the counterfactual variables corresponding to Figure \ref{fig:front1} with intervention on mediator variable $M$.}
\end{minipage}
\caption{DAG and SWIGs of random variables under which average counterfactual outcome can be identified.}
\end{figure}
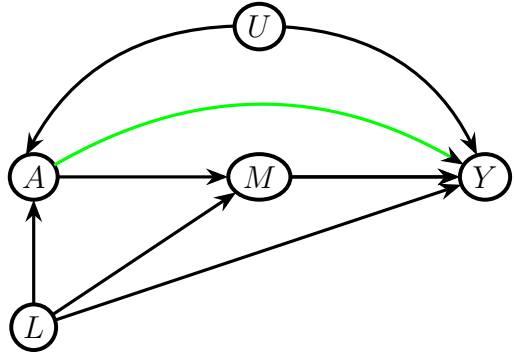
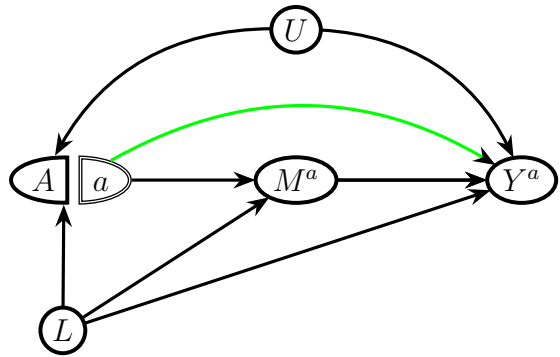
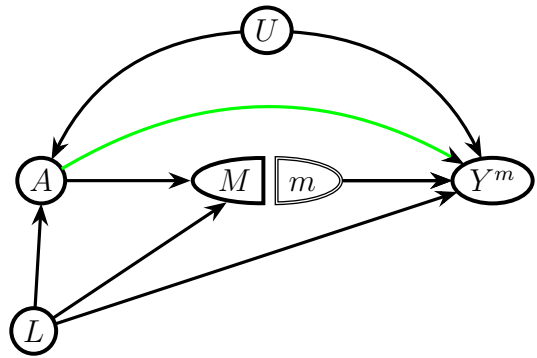

\begin{assumption}[Mediator-Outcome Exchangeability] 
\label{ass:exchMY}
$Y^{a,m} \CI M^a \mid L,A $. 
\end{assumption}
Assumption \ref{ass:noDirect} ensures that $A$ only affects $Y$ through $M$ which holds, for instance, if the green arrow is removed from the graphs in Figure \ref{fig:front1}--\ref{fig:front2b}. Assumption  \ref{ass:exchMY} ensures that the mediator-outcome association is unconfounded which holds, for instance, in the SWIG in Figure \ref{fig:front2b}. Together, Assumptions \ref{ass:pos} and \ref{ass:cons2}--\ref{ass:exchMY} allow unmeasured confounders between exposure and outcome given measured covariates such that $Y^a \nCI A \mid L $ as illustrated in a Directed Acyclic Graph (DAG) in Figure \ref{fig:front1} and SWIGs in Figures \ref{fig:front2a}--\ref{fig:front2b} (see Appendix \ref{sec: app proof frontdoor} for details).


\subsection{Identification of the Population Intervention Indirect Effect}
\label{subsec:purindir}
\citet{Fulcher2020} introduced the PIIE, which they defined as a contrast between the observed mean outcome and the counterfactual outcome if, possibly contrary to fact, the mediator had taken the value that it would have when exposure equals $a^\dagger$. An example of such a contrast is
$E(Y) - E(Y^{M^{a^\dagger},A}).$ \citet{inoue2022causal} studied a closely related estimand, $E(Y^{M^{a^\dagger},A}) - E(Y^{M^{a^\circ},A})$, which they called the path specific frontdoor effect. 

Instead of imposing Assumptions \ref{ass:noDirect}--\ref{ass:exchMY}, \citet{Fulcher2020}  only relied on Assumptions \ref{ass:pos},  \ref{ass:cons2}--\ref{ass:exchAM} and the following cross-world exchangeability assumption to identify the PIIE:
\begin{assumption}[Cross-world exchangeability] 
\label{ass:exchCross}
$Y^{a,m} \CI	M^{a^\dagger} \mid A,  L ~\text{for all values of }a,a^\dagger,m.$
\end{assumption}
Assumption \ref{ass:exchAM} is a statement about the absence of discernible single-world confounding between $M$ and $A$ given $L$, while Assumption \ref{ass:exchCross} is a statement about the absence of indiscernible cross-world confounding between $Y$ and $M$ conditional on $A$ and $L$.

While  the identification strategy proposed by \citet{Fulcher2020} permits the presence of unmeasured common causes of exposure and the outcome, and a direct effect of the exposure on the outcome, their remaining assumptions are not innocuous. First, consistency (Assumption \ref{ass:cons2}) is likely to be violated in our running example; for instance, it is not clear how to imagine well-defined interventions on chronic pain. Second, cross-world independence assumptions like Assumption \ref{ass:exchCross} are controversial because they are untestable even in principle \citep{Robins2010,dawid2000causal}. The untestable cross-world assumption is needed because of the more fundamental fact that the PIIE is a cross-world estimand: a counterfactual quantity that cannot be realized in a single world \citep{richardson2013single}. Because the PIIE can never be directly observed, it is not clear how this estimand can be used as a justification for real-world decisions. This is reflected in our difficulty in finding any correspondence between the PIIE and the substantive questions that were of direct interest to the investigators in our example. More generally, in settings where cross-world estimands often are advocated, other estimands seem to better correspond to questions of practical interest. In particular, \citet{Robins2010} illustrated that the practical relevance of natural (pure) effects tend to be justified by different estimands -- defined by interventions on modified treatments -- using an example on smoking and nicotine. Several other examples have since been given in different settings where other causal effects (e.g.,\ principal stratum) have been advocated in the past \citep{didelez2018causal, stensrud2022conditional}.   



Analogous to \citet{Robins2010}'s interventionist estimands, which often are appropriate when pure (natural) effects are puportedly of interest, our identification result for $E(Y^{a_M})$ can be interpreted as interventionist estimands that are appropriate when the PIIE is puportedly of interest: instead of assuming well-defined interventions on an unmodifiable exposure (Assumption \ref{ass:cons2}) for a parameter only identified under empirically unverifiable cross-world assumptions (Assumption \ref{ass:exchCross}), we consider an intervention on a modifiable exposure $A_M$ that is identifiable under conditions that, in principle, are empirically testable.

 \section{New estimators based on new representations of the efficient influence function of the frontdoor formula}
 \label{sec:estimators}
Because our proposed estimand $E(Y^{a_M})$  is identified by the generalized frontdoor formula, we can apply existing estimators for the generalized frontdoor formula, such as the Augmented Inverse Probability Weighted (AIPW) semiparametric estimator in \citet{Fulcher2020}. However, an issue with the existing semiparametric estimator is that it does not guarantee that the estimate is bounded by the parameter space, which often results in poor performance in finite samples.  In particular, this is a concern our example, where the outcome is a binary indicator of mortality. 
Hence, here we develop new semiparametric estimators that are guaranteed to be bounded by the parameter space. 

Suppose that the observed data $\mathcal{O}=(L,A,M,Y)$ follow a law $P$ which is known to belong to $\mathcal{M}=\{P_\theta:\theta\in \Theta\}$, where $\Theta$ is the parameter space. The efficient influence function $\varphi^{\eff}(\mathcal{O})$ for a causal parameter $\Psi\equiv \Psi(\theta)$ in a {non-parametric model $\mathcal{M}_{\text{np}}$ that imposes no restrictions on the law of $\mathcal{O}$ other than positivity} is given by
${d\Psi(\theta_t)}/{dt}\vert_{t=0} = E\{\varphi^{\eff}(\mathcal{O})S(\mathcal{O})\}$, where ${d\Psi(\theta_t)}/{dt}\mid_{t=0}$ is known as the pathwise derivative of the parameter $\Psi$ along any parametric submodel of the observed data distribution indexed by $t$, and $S(\mathcal{O})$ is the score function of the parametric submodel evaluated at $t=0$ \citep{newey1994,Van2000}.

We will first re-express our causal estimand -- identified by the generalized frontdoor formula -- as a weighted average of two terms given by $\psi_2$ and $\psi_3$: 
\begin{align*}
&\Psi = {P(A=a^\dagger)}\underbrace{E(Y\mid A=a^\dagger)}_{\psi_2} +
P(A=a^\circ)\underbrace{\sum_{m,l}E(Y\mid M=m, L=l, A=a^\circ)f(m\mid a^\dagger,l)f(l\mid a^\circ)}_{\psi_3}.
\end{align*}

Thus, using differentiation rules \citep{Van2000, Ichimura2022,Kennedy2022}, the efficient influence function of $\Psi$ can be realized by finding the efficient influence function of the following parameters: (1) $\psi_1=P(A=a)$, (2) $\psi_2 = E(Y\mid A=a^\dagger) $ and (3) $\psi_3 = \sum_{m,l}E(Y\mid M=m, L=l, A=a^\circ)f(m\mid a^\dagger,l)f(l\mid a^\circ)$. 
We can view $\psi_3$ as equivalent to the identification formula for the $E(Y^{a_M=a^\dagger, a_Y=a^\circ})$ estimand in a separable effect and can, for instance, be identified in an extended DAG of Figure \ref{fig:front1} in the absence of $U$ if the direction of the arrow between $A$ and $L$ is switched (see \citealp{Robins2010,robins2020interventionist,stensrud2021generalized,stensrud2022conditional}). 
The following Theorem allows us to derive semiparametric estimators that are based on an efficient influence function derived from this particular weighted representation of $\Psi$.

\begin{theorem}
The efficient influence function $\varphi^{\text{eff}}(\mathcal{O})$ of the generalized front-door formula in $\mathcal{M}_{\text{np}}$ for $\mathcal{O}=(L,A,M,Y)$ is given by
\begin{align}
\varphi^{\text{eff}}(\mathcal{O}) =& I(A=a^\dagger)Y + I(A=a^\circ)\psi_3 +
\textcolor{blue}{\frac{I(A=a^\circ)P(A=a^\dagger\mid M,L)P(A=a^\circ\mid L)}{P(A=a^\circ\mid M,L)P(A=a^\dagger\mid L)}\{Y-b_0(M,L)\}  +}\nonumber
\\&
\textcolor{blue}{\frac{I(A=a^\dagger)P(A=a^\circ\mid L)}{P(A=a^\dagger\mid L)}\{b_0(M,L) - h_\dagger(L)\} + } \textcolor{blue}{I(A=a^\circ)\{h_\dagger(L) - \psi_3\}} - \Psi \label{eq:eif_gen1}
\end{align}
where in the equation, $b_0(M,L)=E(Y\mid M,L,A=a^\circ)$, $h_\dagger(L) = E(b_0(M,L)\mid A=a^\dagger,L)$ and the Expression in blue is the efficient influence function for $\psi_3$.
The efficient influence function \eqref{eq:eif_gen1} can also be re-expressed as
\begin{align}
\varphi^{\text{eff}}(\mathcal{O}) =& I(A=a^\dagger)Y + I(A=a^\circ)\psi_3 + 
 \frac{I(A=a^\circ)f(M\mid A=a^\dagger,L)}{f(M\mid A=a^\circ,L)}\{Y-b_0(M,L)\}  +\nonumber \\&
 \frac{I(A=a^\dagger)P(A=a^\circ\mid L)}{P(A=a^\dagger\mid L)}\{b_0(M,L) - h_\dagger(L)\} + I(A=a^\circ)\{h_\dagger(L) - \psi_3\} - \Psi,
\label{eq:eif_gen2} 
\end{align}
\end{theorem}

A proof for the efficient influence function can be found in Appendix \ref{sec:EIF_deriv}. 
After some algebra, it can be shown that \eqref{eq:eif_gen2} has an alternative representation given by Equation (5) in Theorem 1 in \citet{Fulcher2020}. 
As such, any regular and asymptotically linear estimator $\hat{\Psi}$ of $\Psi$ in $\mathcal{M}_{\text{np}}$ will satisfy the following property: $\sqrt{n}(\hat{\Psi} - \Psi) = n^{-1/2}\sum_{i=1}^n \varphi^{\text{eff}}(\mathcal{O}_i)+o_p(1)$. 
Furthermore, all regular and asymptotically linear estimators in $\mathcal{M}_{\text{np}}$ are asymptotically equivalent and attain the semiparametric efficiency bound \citep{Bickel1998}.


\subsection{Semiparametric estimators for the generalized frontdoor formula}
Writing the efficient influence function for the generalized frontdoor formula ($\Psi$) given in Expressions \eqref{eq:eif_gen1} or \eqref{eq:eif_gen2} motivates estimators that guarantee sample-boundedness. 
A weighted iterative conditional expectation (Weighted ICE) estimator that guarantees sample-boundedness is presented in the following algorithm. In what follows, we let $\mathbb{P}_n(X) = n^{-1}\sum_{i=1}^n X_i$ and let $g^{-1}$ denote a known inverse link function satisfying $\inf(\mathbf{Y})\leq g^{-1}(u)\leq \sup(\mathbf{Y})$, for all $u$, where $\mathbf{Y}$ is the sample space of $Y$ (e.g., a logit link for dichotomous $Y$).

\begin{breakablealgorithm}
\renewcommand{\theenumi}{\Alph{enumi}}   
\caption{Algorithm for Weighted ICE (generalized frontdoor formula)}          
\begin{algorithmic} [1]      
\item Non-parametrically compute $P(A=a^\circ)$ and $P(A=a^\dagger)$.
\item Compute the maximum likelihood estimate (MLE) $\hat{\kappa}$ of ${\kappa}$ from the observed data for the exposure or treatment model $P(A=a \mid L;\kappa)$. In addition, compute the MLE $\hat{\alpha}$ of ${\alpha}$ from the observed data for the exposure or treatment model $P(A=a \mid M,L;\alpha)$, or compute the MLE $\hat{\gamma}$ of $\gamma$ from the observed data for the mediator model $P(M=m \mid A, L;\gamma).$
\item In the individuals whose $A=a^\circ$, fit a regression model $Q({M,L};\theta)= g^{-1}\{\theta^T\phi(M,L)\}$ for $b_0(M,L)=E(Y \mid M,L,A=a^\circ)$ where the score function for each observation is weighted by $\hat{W}_1$. Here, $\hat{W}_1$ equals $$\frac{P(A=a^\circ\mid L;\hat{\kappa})P(A=a^\dagger\mid M,L;\hat{\alpha})}{P(A=a^\dagger\mid L;\hat{\kappa})P(A=a^\circ\mid M,L;\hat{\alpha})}$$ if $\hat{\alpha}$ was estimated in the previous step, or $\hat{W}_1$ equals $$\frac{f(M\mid A=a^\dagger, L;\hat{\gamma})}{f(M\mid A=a^\circ, L;\hat{\gamma})}$$
if $\hat{\gamma}$ was estimated in the previous step, and $\phi(M,L)$ is a known function of $M$ and $L$.
More specifically, we solve for $\theta$ in the following estimating equations:
\begin{equation*}
   \mathbb{P}_n \left[ I(A=a^\circ)\phi(M,L)\hat{W}_1 \left\{Y - Q({M,L};\theta)\right\}\right] = 0.
\end{equation*}
\item In those whose $A=a^\dagger$, fit a model $R(L;\eta)=g^{-1}\{\eta^T\Gamma(L)\}$ for $h_\dagger(L)=E(b_0(M,L) \mid L,A=a^\dagger)$ where the score function for each observation is weighted by $$\hat{W}_2 = \frac{P(A=a^\circ\mid L;\hat{\kappa})}{P(A=a^\dagger\mid L; \hat{\kappa})}.$$
Here, $\Gamma(L)$ is a known function of $L$.
More specifically, we solve for $\eta$ in the following estimating equations:
\begin{equation*}
   \mathbb{P}_n \left[ I(A=a^\dagger)\Gamma(L)\hat{W}_2\left\{Q({M,L};\hat{\theta}) - R(L;\eta)\right\}\right] = 0.
\end{equation*}
\item In those whose $A=a^\circ$, fit an intercept-only model $T(\beta)= g^{-1}(\beta)$ for $\psi_3 = E\{h_\dagger(L)\mid A=a^\circ\}$. More specifically, we solve for $\beta$ in the following estimating equations:
\begin{equation*}
   \mathbb{P}_n \left[ I(A=a^\circ)\left\{R({L};\hat{\eta}) - T(\beta)\right\}\right] = 0.
\end{equation*}
\item Compute $\hat{T} \equiv T(\hat{\beta})$ for all observations.
\item Estimate $\hat{\Psi}_{WICE}=\mathbb{P}_n\{I(A=a^\dagger)Y + I(A=a^\circ)\hat{T}\}.$
\end{algorithmic}
\end{breakablealgorithm}

In Algorithm 1, steps 3, 4 and 5 ensure that the estimates for $\psi_3 = E\{h_\dagger(L)\mid A=a^\circ\}$ are sample bounded. Moreover, in Step 6 it is clear that $\hat{\Psi}_{WICE}$ is a convex combination of $Y$ and estimates for $\psi_3$, both of which are bounded by the range of the outcome $Y$. Thus, $\hat{\Psi}_{WICE}$ will also be sample-bounded. 
In Appendix \ref{sec:appendix_robust}, we prove that the proposed estimator, which is based on the efficient influence function given by \eqref{eq:eif_gen1} and \eqref{eq:eif_gen2}, is robust against 3 classes of model misspecification scenarios. 

\begin{theorem}
Under standard regularity conditions, the weighted ICE estimator $\hat{\Psi}_{WICE}$ where a model for $P(A=a\mid M,L)$ is specified will be consistent and asymptotically normal under the union model $\mathcal{M}_{union}=\mathcal{M}_{1}\cup \mathcal{M}_2\cup \mathcal{M}_3$ where we define:
\begin{enumerate}[leftmargin=*]
    \item Model $\mathcal{M}_1$: working models for $P(A=a\mid M,L)$ {and} $P(A=a\mid L)$ are correctly specified.
    \item Model $\mathcal{M}_2$: working models for $b_0(M,L)$ and $h_\dagger(L)$ are correctly specified.
    \item Model $\mathcal{M}_3$: working models for $b_0(M,L)$ and $P(A=a\mid L)$ are correctly specified.
\end{enumerate}
The weighted ICE estimator $\hat{\Psi}_{WICE}$ where a model for $P(M=m\mid A,L)$ is specified will be consistent and asymptotically normal under the union model $\mathcal{M}_{union}=\mathcal{M}_{1}\cup \mathcal{M}_2\cup \mathcal{M}_3$ where we define (1) Model $\mathcal{M}_1$: working models for $P(M=m\mid A,L)$ {and} $P(A=a\mid L)$ are correctly specified; (2) Model $\mathcal{M}_2$: working models for $b_0(M,L)$ and $h_\dagger(L)$ are correctly specified; (3) Model $\mathcal{M}_3$: working models for $b_0(M,L)$ and $P(A=a\mid L)$ are correctly specified.
Moreover, $\hat{\Psi}_{WICE}$ is locally efficient in the sense that it achieves the semiparametric efficiency bound for $\Psi$ in $\mathcal{M}_{np}$, i.e., $E\left[\varphi^{\text{eff}}(\mathcal{O})^2  \right]$, at the intersection model given by $\mathcal{M}_{intersection} = \mathcal{M}_{1}\cap \mathcal{M}_2\cap \mathcal{M}_3$.
\end{theorem}
Note that it is possible that the weighted ICE estimator is consistent when the models for $b_0(M,L)$ and $E(M\mid A,L)$ are correctly specified. For instance, this would be the case when $Y$ and $M$ are continuous, in which case the model $R(L;\eta)$ for $h_\dagger(L)$ can also be correctly specified in this model specification scenario. 

Unlike the estimator proposed by \citet{Fulcher2020}, the estimator proposed here requires specification of four models instead of three, and thus our proposed estimator offers a different robustness property against model misspecification compared to that of the AIPW estimator in \citet{Fulcher2020}. Consequently, the AIPW estimator in \citet{Fulcher2020}, which requires specification of only three (vs. four) working models, is robust against two (vs. three) classes of model misspecification scenarios.  Specifically, it will be consistent when at least (1) the models for $b_0(M,L)$ and $P(A=a\mid L)$ are correctly specified, \textit{or} (2) only the model for $P(M=m\mid A,L)$ is correctly specified. 
In Appendix \ref{sec:appendix_estimators}, we describe an iterative algorithm that preserves the double robustness property of \citet{Fulcher2020} for binary mediators. However, we do not pursue this iterative algorithm in our main manuscript as it is known to be more computationally intensive and can run into convergence issues (see \citealp{van2009,Van2018targeted}).

\sloppy
We view the extra model specification as a trade-off between robustness and sample-boundedness. Moreover, when (i) $M$ is a continuous variable, as in the `Safer deliveries' application in \citet{Fulcher2020}, and/or (ii) there are multiple mediator variables ($M_1$, $M_2$, $M_3$\ldots), the model for the mediator variable is difficult to correctly specify, and thus our proposed estimator in this case will be more useful. 
In Appendix \ref{sec:appendix_robust}, we prove that in the absence of $L$ our weighted ICE estimator based on the efficient influence function for the (non-generalized) frontdoor formula is doubly robust in the sense that it will be consistent as long as the model for $P(A=a\mid M)$ -- or $P(M=m\mid A)$, depending on the representation -- or the model for $E(Y\mid A=a^\circ, M)$ is correctly specified. 
As such, for the (non-generalized) frontdoor formula in the absence of $L$, the double robustness property of our estimator is the same as that of \citet{Fulcher2020}. 

We also describe a targeted maximum likelihood estimator (TMLE) for $\Psi$, which is a variation of the weighted ICE estimator given above. The TMLE can accommodate complex machine learning algorithms for estimating all nuisance functions \citep{Van2011, Chernozhukov2018, Wen2021}.

\newpage
\begin{breakablealgorithm}
\renewcommand{\theenumi}{\Alph{enumi}}   
\caption{Algorithm for Targeted maximum likelihood (generalized frontdoor formula)} 
\begin{algorithmic} [1]      
\item Non-parametrically compute $P(A=a^\circ)$ and $P(A=a^\dagger)$.
\item Obtain estimates $\hat P(A=a \mid L)$, $\hat P(A=a \mid M,L)$ and $\hat P(M=m \mid A, L)$ of $ P(A=a \mid L)$, $P(A=a \mid M,L)$ (or $P(M=m \mid A, L)$), respectively, possibly using machine learning methods.
\item 
\begin{enumerate}
    \item In the individuals whose $A=a^\circ$, compute $\hat{Q}({M,L})$ by regressing $Y$ on $(M,L)$. Here, $\hat{Q}({M,L})$ is possibly estimated using machine learning methods, and it denotes an initial estimate for $b_0(M,L)=E(Y \mid M,L,A=a^\circ)$.
    \item Update the previous regression. Specifically, in the individuals whose $A=a^\circ$, fit a intercept-only regression model $Q^\ast({M,L};\delta)=  g^{-1}[g\{\hat{Q}({M,L})\}+\delta]$ where the score function for each observation is weighted by $\hat{W}_1$ (defined previously). More specifically, we solve for $\delta$ in the following estimating equations:
\begin{equation*}
   \mathbb{P}_n \left[ I(A=a^\circ)\hat{W}_1 \left\{Y - Q^\ast({M,L};\delta)\right\}\right] = 0
\end{equation*}
\end{enumerate}
\item \begin{enumerate}
    \item In those whose $A=a^\dagger$, compute $\hat{R}(L)$ by regressing $Q^\ast({M,L};\hat{\delta})$ (as outcome, obtained from last step) on $L$. Here, $\hat{R}(L)$ is possibly estimated using machine learning methods, and it denote an initial estimate of $h_\dagger(L)=E(b_0(M,L) \mid L,A=a^\dagger)$.
    \item Update the previous regression. Specifically, in those whose $A=a^\dagger$, fit an intercept-only regression model $R^\ast(L;\nu)= g^{-1[}\{g\{\hat{R}(L)\}+\nu]$ where the score function for each observation is weighted by $\hat{W}_2$.
More specifically, we solve for $\nu$ in the following estimating equations:
\begin{equation*}
   \mathbb{P}_n \left[ I(A=a^\dagger)\hat{W}_2\left\{Q^\ast({M,L; \hat{\delta}}) - R^\ast(L;\nu)\right\}\right] = 0
\end{equation*}
\end{enumerate}
\item In those whose $A=a^\circ$, fit another regression model $T(\beta)= g^{-1}(\beta)$ for $\psi_3 = E\{h_\dagger(L)\mid A=a^\circ\}$ with just an intercept. More specifically, we solve for $\beta$ in the following estimating equations:
\begin{equation*}
   \mathbb{P}_n \left[ I(A=a^\circ)\left\{R^\ast({L};\hat{\nu}) - T(\beta)\right\}\right] = 0
\end{equation*}
\item Compute $\hat{T} \equiv T(\hat{\beta})$ for all observations.
\item Estimate $\hat{\Psi}_{TMLE}=\mathbb{P}_n\{I(A=a^\dagger)Y + I(A=a^\circ)\hat{T}\}$
\end{algorithmic}
\end{breakablealgorithm}


\newcommand\norm[1]{\lVert#1\rVert}
\newcommand\normx[1]{\Vert#1\Vert}
\begin{theorem}[Weak convergence of TMLE]
    Suppose that the conditions given in Appendix \ref{sec:asymp} hold, and further suppose that the following condition also holds:
    \begin{align*}
        \norm{\hat h_\dagger(L)-h_\dagger(L)}~ \norm{\hat f(A\mid L) - f(A\mid L)} + &\norm{\hat b_0(M,L)-b_0(M,L)}~ \norm{\hat f(A\mid L) - f(A\mid L)} +
        \\& \norm{\hat b_0(M,L)-b_0(M,L)}~ \norm{\hat f(A\mid M, L) - f(A\mid M, L)} = o_p(n^{-1/2}),
    \end{align*}
    where $\norm{f(x)}_2 = \left\{\int |f(x)|^2dP(x)\right\}^{1/2}$, i.e. the $L_2(P)$ norm.
Then, $$\sqrt{n}\left(\widehat{\Psi}_{{TMLE}} - \Psi\right) \rightsquigarrow N(0,\sigma^2),~~\text{where}~\sigma^2 = \text{Var}(\varphi^{\text{eff}}(\mathcal{O})).$$ The variance of TMLE can be estimated using the sandwich variance estimator (where nuisance function estimators are the same as those in TMLE), or via bootstrap.
\label{thm:three}
\end{theorem}

The advantage of using TMLE (or the AIPW estimator of \citealp{Fulcher2020}) with machine learning algorithms for the nuisance functions is that the estimator is still consistent and asymptotically normal as long as the nuisance functions converge to the truth at rates faster than $n^{-1/4}$ \citep{Robins2008,Chernozhukov2018} (see Appendix \ref{sec:asymp}). Nevertheless, in real world applications there is no guarantee that such rates of convergence can be attained when more flexible algorithms such as neural network or random forest are used. Moreover, there is no guarantee that these machine learning methods will exhibit more or less bias compared with parametric models \citep{Liu2020}. 

\section{Simulation study}
\label{sec:sims}
We conducted a simulation study to demonstrate that (1) our estimand, like the PIIE, is robust to unmeasured confounding between exposure and outcome, (2) our proposed estimator is more robust to model misspecification compared with estimators such as the Inverse Probability Weighted (IPW) estimator and the ICE estimator (described in Appendix \ref{sec:appendix_estimators}), and (3) unlike the AIPW estimator of \citet{Fulcher2020} our estimator can also ensure sample boundedness in any finite sample size setting.

The simulation study was based on 1000 simulated data sets of sample sizes $n=100,~250$ and $500$. We compared the bias, standardized bias and efficiency of IPW, ICE, AIPW and weighted ICE estimators. Standardized bias is $100\times$[(Average Estimate$-$True Parameter)/empirical standard deviation of the parameter estimates]. It has been suggested that for $n=500$, anything greater than an absolute standardized bias of approximately 40\% will have a `noticeable adverse impact on efficiency, coverage, and error rates' \citep{Collins2001}. The true value of $\Psi=0.0144$ was calculated by generating a Monte Carlo sample of size $10^7$. We intentionally chose a rare outcome to compare the performance of the estimators where the AIPW estimator had a non-trivial chance of falling outside of the parameter space.

The data-generating mechanism for our simulations and model specifications are provided in Table \ref{table:DGM}. We consider four scenarios to illustrate the robustness of our proposed estimator to model misspecification: (1) all models are approximately correctly specified, (2) only the models for $b_0(M,L)$ and $h_\dagger(L)$ are approximately correctly specified,  (3) only the models for $b_0(M,L)$ and $P(A=a\mid L)$ are approximately correctly specified, and (4) only the model for $P(M=m\mid A,L)$ is correctly specified \textit{and} the model for $P(A=a\mid L)$ is approximately correctly specified. The correct mediator model in the specification scenarios is the one used in the data generation process, and the exposure and outcome models are approximately correctly specified by including pairwise interactions between all the variables to ensure flexibility.

Table \ref{table:simresult} shows the results from the simulation study. Consistent with our theoretical derivations, when all of the working models are (approximately) correctly specified, all of the estimators become nearly unbiased as the sample size increases. For $n=500$, the AIPW estimator and our proposed weighted ICE estimator are also nearly unbiased in the three model misspecification settings whereas the IPW and ICE estimators are not all unbiased.
In Appendix \ref{sec:extrasim}, we show an additional simulation study where the variables are all binary and thus the correct exposure and outcome models are saturated models that cannot be misspecified. The results further show the robustness of our estimator in model misspecification scenarios.

Finally, ICE, AIPW and weighted ICE have comparable standard errors when all of the models are correctly specified, but all three had lower standard errors compared with IPW. However, our results also show that AIPW estimator has poorer finite sample performance when $n=100$ compared with all other estimators. Moreover, for $n=100$, there were 90, 59 and 88 simulated data sets where estimates fell below 0 in scenarios 1, 3 and 4, respectively.

\begin{table}[h!]
\centering
\begin{tabular}{||p{3cm} p{12cm} ||} 
 \hline \hline
 \multicolumn{2}{||c||}{Data generating mechanism} \\ [0.5ex] 
 \hline
  $U\sim$ & $\text{Ber}(0.5)$  \\ 
  $L_1\sim$ & $\text{Normal}(0,1)$  \\
  $L_2\sim$ & $\text{Ber}\{\expit(1+2L_1)\}$  \\ 
  $A\sim$ & $\text{Ber}\{\expit(-1-3L_1+L_2+5L_1L_2+2U)\}$  \\ 
  $M\sim$ & $\text{Ber}\{\expit(1-A-2L_1+2L_2+3L_1L_2)\}$  \\ 
  $Y\sim$ & $\text{Ber}\{\expit(-4+2A+M-2AM+2L_1-2L_2-5L_1L_2-U)\}$  \\ 
   \hline
 \multicolumn{2}{||c||}{Model misspecification} \\ [0.5ex]
  \hline
Scenario 2 & $P(M=m\mid L,A; \gamma) = \expit(\gamma_0 + \gamma_1 A + \gamma_2 L_1 + \gamma_3 L_2)$\\
 & $P(A=a\mid L; \kappa) = \expit(\kappa_0 + \kappa_1 L_1 + \kappa_2 L_1^2)$\\ [1ex]
Scenario 3 & $P(M=m\mid L,A; \gamma) = \expit(\gamma_0 + \gamma_1 A + \gamma_2 L_2)$\\ 
 & $R(L;\theta) = \expit(\eta_0 + \eta_1 L_2)$\\ [1ex]
Scenario 4 & $Q(M,L;\theta) = \expit(\theta_0 + \theta_1 M + \theta_2 L_1 + \theta_3 L_2)$\\
 & $R(L;\theta) = \expit(\eta_0 + \eta_1 L_2 (1-L_1))$\\
 \hline\hline
\end{tabular}
\caption{Data generating mechanism and model misspecifications for scenarios in simulation study.}
\label{table:DGM}
\end{table}

\begin{table}
\centering
\small
\begin{tabular}{ |l | c c c | c c c| c c c | c c c|  }
\hline
& \multicolumn{3}{|c|}{Scenario 1} & \multicolumn{3}{|c|}{Scenario 2}  & \multicolumn{3}{|c|}{Scenario 3}  & \multicolumn{3}{|c|}{Scenario 4} \\
\hline
& Bias  & SE & Bias$_{s}$ & Bias  & SE & Bias$_{s}$ & Bias  & SE & Bias$_{s}$ & Bias  & SE & Bias$_{s}$ \\
\hline
\multicolumn{13}{|c|}{$n=100$}  \\
\hline
IPW	&0.21	&2.19	&9.52	&-0.23	&1.38	&-16.67	&0.21	&2.19	&9.52	&3.13	&2.12	&147.8\\
ICE	&0.15	&1.74	&8.42	&0.15	&1.74	&8.42	&-0.28	&1.36	&-20.7	&0.55	&1.93	&28.53\\
AIPW	&0.49	&1.94	&25.03	&-	&-	&-	&0.61	&2.19	&27.71	&0.31	&1.72	&18.31\\
WICE	&0.20	&1.81	&10.82	&0.15	&1.75	&8.57	&0.17	&1.96	&8.551	&0.23	&1.72	&13.52\\
\hline
\multicolumn{13}{|c|}{$n=250$}  \\
\hline
IPW	&0.12	&1.33	&9.05	&-0.33	&0.86	&-38.04	&0.12	&1.33	&9.05	&3.45	&1.33	&259.62\\
ICE	&0.09	&1.05	&8.97	&0.09	&1.05	&8.97	&-0.36	&0.92	&-38.93	&0.39	&1.13	&34.17\\
AIPW	&0.16	&1.07	&14.89	&-	&-	&-	&0.16	&1.18	&13.16	&0.07	&0.92	&7.26\\
WICE	&0.09	&1.02	&9.06	&0.08	&1.03	&7.92	&0.09	&1.22	&7.38	&0.01	&0.88	&1.26\\
\hline
\multicolumn{13}{|c|}{$n=500$}  \\
\hline
IPW	&0.07	&0.94	&7.75	&-0.40	&0.54	&-74.8	&0.07	&0.94	&7.75	&3.60	&0.98	&365.37\\
ICE	&0.04	&0.68	&6.34	&0.04	&0.68	&6.34	&-0.45	&0.54	&-83.93	&0.36	&0.72	&49.71\\
AIPW	&0.04	&0.68	&5.73	&-	&-	&-	&0.05	&0.88	&5.72	&0.03	&0.57	&4.89\\
WICE	&0.05	&0.68	&6.96	&0.04	&0.67	&6.51	&0.04	&0.83	&5.33	&0.01	&0.55	&1.02\\
\hline
\end{tabular}
\caption{Simulation results: Bias, standard error (SE), and standardized bias (Bias$_s$) are multiplied by 100, and true value was $\Psi=0.0144$. For $n=100$, 90, 59 and 88 simulated data sets had estimates that fell below 0 in scenarios 1, 3 and 4, respectively. For $n=250$, 16, 38 and 14 simulated data sets had estimates that fell below 0 in scenarios 1, 3 and 4, respectively. For $n=500$, 5, 31, and 1 simulated data set(s) had estimates that fell below 0 in scenarios 1, 3 and 4, respectively.}
\label{table:simresult}
\end{table}

\section{Application}
\label{sec:application}
Motivated by the chronic pain example, we applied our results to study the causal effect of modified prescription policies for opioids on mortality in patients with chronic pain. The intervention of interest is a new prescription policy, in which the doctor is set to ignore (i.e., consider as absent, possibly counter to fact) each patient's chronic pain for the purpose of opioid prescription decisions, and otherwise use measured covariates as usual (according to standard treatment guidelines).  Thus, this policy is an intervention on a \textit{doctor's perception of chronic pain} $A_M$. This intervention is consistent with a recently proposed policy by the CDC that implores practicing physicians to no longer consider chronic pain as an indication for opioid therapy.
While the doctors' perceptions are not explicitly measured in the observed data, we may reasonably assume that (in the absence of intervention) such perceptions perfectly correspond with the medical reality of the patient (that is, $A=A_M$ almost surely).  

We used the dataset from \citet{inoue2022causal}, which includes observations from the US NHANES that are linked to a national mortality database (National Death Index). The NHANES study consists of a series of in-depth in-person interviews, medical and physical examinations, and laboratory tests aimed at understanding various emerging needs in public health and nutrition. Sample data are from 1999--2004 and include information on individuals' chronic pain statuses ($A$), opioid prescriptions ($M$), mortality ($Y$), and covariates ($L$) including age, sex assigned at birth (male and female), race (non-Hispanic White, non-Hispanic Black, Mexican-American, or others), education levels (less than high school, high school or General Education Degree, or more than high school), poverty-income ratio (the ratio of household income to the poverty threshold), health insurance coverage, marital status, smoking, alcohol intake, and anti-depressant medication prescription. Let $A_M$ denote the \textit{intervening} variable: the doctor's perception of the chronic pain of the patient. 

Our sample included 12037 individuals. Following \citet{inoue2022causal}, an individual is considered to have chronic pain if they reported pain for at least three months by the International Association for the Study of Pain criteria \citep{1994classification}.  Moreover, data on prescription medications for pain relief used in the past 30 days were collected in the in-person interview. Opioids identified through the process include codeine, fentanyl, oxycodone, pentazocine and morphine. About 16\% of the individuals in the sample experienced chronic pain and approximately 5\% of the individuals in the sample reported using opioids. Detailed data description and summary statistics can be found in \citet{inoue2022causal}. 

We estimated the cumulative incidence $E(Y^{a_M=0})$ and the causal contrast $E(Y) - E(Y^{a_M=0})$ using ICE, IPW and our weighted ICE estimator by specifying logistic regression models for the outcome, mediator and exposure. We used the same logistic regression models as that of \citet{inoue2022causal} by adjusting for all the previously-listed measured covariates as potential confounders. All 95\% confidence intervals were based on the 2.5 and 97.5 percentiles of a non-parametric bootstrap procedure with 1000 bootstrapped samples.  



\sloppy
The ICE procedure estimated the $E(Y^{a_M=0})$ to be 4.76\% (95\% CI$=(4.36,~5.16)$) in three years and 8.55\% (95\% CI$=(8.04,~9.09)$) in five years. 
The IPW procedure estimated this cumulative incidence to be 4.95\% (95\% CI$=(4.57,~5.35)$) in three years and 8.82\% (95\% CI$=(8.30,~9.35)$) in five years.
The weighted ICE procedure estimated this cumulative incidence to be 4.76\% (95\% CI$=(4.36,~5.16)$) in three years and 8.55\% (95\% CI$=(8.04,~9.09)$) in five years.

Moreover, the ICE procedure estimated $E(Y) - E(Y^{a_M=0})$ to be 0.22\% (95\% CI$=(-0.32,~0.82)$) in three years and 0.25\% (95\% CI$=(0.09,~0.40)$) in five years. 
The IPW procedure estimated this causal contrast to be 0.02\% (95\% CI$=(-0.38,~0.40)$) in three years and -0.03\% (95\% CI$=(-0.55,~0.49)$) in five years.
Finally, the weighted ICE procedure estimated this causal contrast to be 0.22\% (95\% CI$=(-0.33,~0.82)$) in three years and 0.25\% (95\% CI$=(0.09, 0.40)$) in five years.

The three estimators produced similar results. We also applied the TMLE estimator described herein where the nuisance functions are estimated using the Super Learner ensemble (library of candidates including generalized additive models and multivariate adaptive regression Splines) and found similar results. For instance, the TMLE procedure estimated $E(Y) - E(Y^{a_M=0})$ to be 0.24\% (95\% CI$=(0.08, 0.39)$) in five years.
As such, our analysis suggests that under an intervention on the doctor's perception of chronic pain when making decisions about opioid treatment, the cumulative incidence of death is almost identical to the cumulative incidence in the observed data after three years, but may decrease risk very slightly after five years.

\section{Discussion}
\label{sec:discussion}
We have derived identification results that justify the use of the frontdoor formula in new settings, reflecting questions of practical interest, where unmeasured confounding is a serious concern. 
Our identification results do not rely on ill-defined interventions or cross-world assumptions. 
Specifically, we proposed an estimand defined by an intervention on a modifiable descendant of an exposure or treatment, which we call an \textit{intervening} variable.  
Like the previously proposed PIIE, our proposed estimand is identified by the frontdoor formula even in the presence of a direct effect of the exposure on the outcome not mediated by an intermediate variable. 
But unlike the PIIE, our estimand is identifiable under conditions that, in principle, are empirically testable.
In addition, we presented an example in which our proposed estimand is practically relevant. In this example, the exposure variable -- chronic pain -- was difficult, if not impossible, to intervene on. However, we argued that interventions on the {intervening} variable, rather than the exposure, are often of practical interest in settings where interventions on exposures are ill-defined.




When our proposed estimand is identified by the frontdoor formula in the absence of $L$, our proposed estimator and the existing AIPW estimator of \citet{Fulcher2020} are both doubly robust in the sense that they are consistent as long as (1) the model for $P(A=a\mid M)$ (or $P(M=m\mid A)$) is correctly specified or (2) the model for $b_0(M)$ is correctly specified. 
For the generalized frontdoor formula, our proposed estimator is triply robust when (1) the models for $P(A=a\mid M,L)$ (or $P(M=m\mid A,L)$) \textit{and} $P(A=a\mid L)$ are correctly specified, (2) the models for $b_0(M,L)$ and $h_\dagger(L)$ are correctly specified, or (3) the models for $b_0(M,L)$ and $P(A=a\mid L)$ are correctly specified. Compared with existing AIPW estimator of \citet{Fulcher2020}, there is one disadvantage of our estimator for the generalized g-formula; our estimator requires specification of four models instead of three in the AIPW estimator. As a result, the AIPW estimator is doubly robust when (1) the models for $b_0(M,L)$ and $P(A=a\mid L)$ are correctly specified, \textit{or} (2) only the model for $P(M=m\mid A,L)$ is correctly specified. 
However, we view this as a trade-off to guarantee sample-boundedness, which is a desirable property in estimators based on the efficient influence function.
Indeed, as stated in \citet{Robins2007}, ``With highly variable weights, `boundedness' trumps unbiasedness."
A key advantage of our semiparametric estimator is that estimates of $\Psi$ are ensured to be bounded by the parameter space, regardless of the sample size and variability of inverse probability weights. 

In practice it is advised to define richly parameterized models for $P(A=a\mid L)$ and $h_\dagger(L)$ to ameliorate model incompatibility issues between $P(A=a\mid L)$ and $P(A=a\mid M, L)$ and between $b_0(M,L)$ and $h_\dagger(L)$ \citep{Vansteelandt2007,Tchetgen2014}.
However, similar to the AIPW estimator in \citet{Fulcher2020}, our proposed estimator can also accommodate machine learning algorithms through trivial modifications, which in principle can achieve $\sqrt{n}$-consistency as long as the nuisance functions converge at sufficiently fast rates. 
Moreover, when the mediator variable is continuous, the AIPW estimator of \citet{Fulcher2020} involves a preliminary estimator of a density ratio. Direct estimation of the density ratio is often cumbersome as correct specification of the probability density of $M$ is difficult. This is particularly challenging task when data-adaptive estimators are used for estimating high-dimensional nuisance parameters.
Our proposal allows for an alternative estimation procedure to accommodate continuous mediator variables by modeling the propensity score of exposure/treatment instead. 
Finally, our semiparametric estimator can be applied whenever the frontdoor formula identifies the parameter of interest, which e.g.,\ could be the ACE, PIIE or our interventionist estimand.
Our results also motivate future methodological work. In particular, we aim to generalize our results to longitudinal settings, involving time-varying treatments.

\section*{Acknowledgement}
The authors thank Dr. Kosuke Inoue for the access to the NHANES dataset used in the data application. Lan Wen is supported by the Natural Sciences and Engineering Research Council of Canada (NSERC) Discovery Grant [RGPIN-2023-03641]. Mats J.\ Stensrud and Aaron L.\ Sarvet are supported by the Swiss National Science Foundation, grant $200021\_207436$.

\bibliographystyle{apalike}
\bibliography{refs}

\end{document}


\title[Web Appendix: Intervening variables and unmeasured confounding]{Web Appendix for ``Causal effects of intervening variables in settings with unmeasured confounding"}
\author{}
\maketitle

\appendix 

\section{Example from \citet{Fulcher2020}}
\label{eg:fulcher}

\begin{example}[The safer deliveries program, \citep{Fulcher2020}]
\normalfont 
\label{ex:fulcher}
The `Safer deliveries' program was designed to reduce the relatively high rates of maternal and neonatal mortality in Zanzibar, Tanzania. The program provided counselling to pregnant women preparing for delivery. Women deemed to be in a high pregnancy ``risk category'',  based on a mobile device algorithm, were instructed to deliver at a referral hospital, a specialty healthcare resource that generally incurred higher expenses for the women's family. Then, given the recommended delivery location, the mobile device algorithm also calculated an amount that they recommended the family should save in anticipation of future obstetric care costs (i.e. a ``tailored savings recommendation''). At a later point in the study, the amount that the families actually saved for this purpose was recorded in the data (i.e. the ``actual savings''). 

Using data from the `Safer deliveries' program,  \cite{Fulcher2020} aimed to evaluate ``the effectiveness of this tailored savings recommendation by risk category on actual savings". They reported estimates of the PIIE of delivery risk (high risk versus low/medium risk exposure) on actual savings at delivery (outcome), mediated by a recommended savings amount calculated by the mobile device algorithm. As noted in \cite{Fulcher2020}, there may be unmeasured confounding between a participant's recommended risk category and actual savings at delivery, for example by socioeconomic factors and individual's health-seeking behaviour. 
\end{example}

\citet{Fulcher2020} argued that the PIIEs in Example A.\ref{ex:fulcher} was an appropriate estimand to study ``the effectiveness of this tailored savings recommendation" for pregnant women. However, it is not clear that the plain English justification translates to a PIIE defined by interventions on a woman's recommended risk category, an exposure that is non-interveneable or of limited scientific interest. To interpret the results of their analysis we either have to consider:

\begin{enumerate}
      \item obstetric risk category to be \textit{defined} as a composite of various embodied socio-demographic and clinical features, in which case intervention on obstetric risk category can only be defined as interventions on the constituent components used to characterize the exposure. Such an intervention would be difficult to imagine as all of these embodied socio-demographic and clinical features may be hard to identify; or
      \item obstetric risk category to be \textit{defined} as a conceptually distinct feature from the measured socio-demographic and clinical features used to compute it, and thus possibly manipulable separately from these features. In this case the ``risk category'' variable would simply be no more or less than the computed ``risk category'' that appears on the screens of mobile devices, and these risk categories could be manipulated simply by intervening on the software run on these devices. However, such interventions would be of a substantively different nature with profound differences in interpretation and will have different implications for policy-makers.
      Moreover, the exposure ``risk category'' would not be susceptible to unmeasured confounding.
  \end{enumerate}

Considering effects of intervening variables ameliorates this ambiguity and also clarifies assumptions. Specifically, an intervention that avoids these difficulties would be to fix the \textit{output} value from the algorithm, so that it recommends a delivery location as usual, but the patient's recommended savings amount is based on the delivery location the original algorithm would recommend if that patient had been deemed to be at low risk for obstetric complications. We can explicitly define the algorithm's computed risk category as $A_M$ (a modifiable \textit{intervening} variable) that is distinct from the patient's non-modifiable embodied risk category ($A$). In the observed data, $A = A_M$ with probability 1. However, we could conceive an intervention that modifies this intervening variable $A_M$ without changing the exposure $A$.


\newpage
\section{Proofs of identification the Frontdoor Formula}
\label{sec: app proof frontdoor}
The proof for the frontdoor identifying formula \eqref{eq: simple frontdoor formula} is given as follows:
    \begin{align*}
        E(Y^{a^\dagger}) &= \sum_m E(Y^{a^\dagger} \mid M^{a^\dagger} = m)P(M^{a^\dagger} = m)
        \\&  = \sum_m E(Y^{a^\dagger} \mid M^{a^\dagger} = m)P(M = m\mid A=a^\dagger)~~\text{(By A5, A6)}
        \\& = \sum_m E(Y^{a^\dagger,m} \mid M^{a^\dagger} = m)p(m\mid a^\dagger)~~\text{(By A6)}
        \\& = \sum_{a,m}E(Y^{a^\dagger,m}\mid A=a, M^{a^\dagger}=m)p(a)p(m\mid a^\dagger)~~\text{(By A5)}
         \\& = \sum_m p(m\mid a^\dagger) \sum_{a}E(Y\mid A=a, M=m)p(a)~~\text{(By A6, A7, A8)}
    \end{align*}
Alternatively, we also provide a slightly different proof:
\begin{align*}
E(Y^{a^\dagger}) &= \sum_u E(Y^{a^\dagger}\mid U=u)f(u)
\\ &  = \sum_u E(Y^{a^\dagger}\mid A=a^\dagger, U=u)f(u)
\\ &  = \sum_u E(Y\mid A=a^\dagger, U=u)f(u)
\\ &  = \sum_{u,m} E(Y\mid A=a^\dagger, U=u, M=m)f(m\mid a^\dagger, u)f(u)
\\ &  = \sum_m f(m\mid a^\dagger) {\sum_u E(Y\mid A=a^\dagger, U=u, M=m)f(u)}
\\ &  = \sum_m f(m\mid a^\dagger) \underbrace{\sum_u E(Y\mid U=u, M=m)f(u)}_{(\ast)}
\\ &  = \sum_{m} f(m\mid a^\dagger) \sum_a E(Y\mid A=a, M=m)f(a).
\end{align*}
Aside from probability laws, we note the following conditions that are used in the proof above: line 2 follows by conditional exchangeability of $Y^{a^\dagger}$ and $A$ conditional on $U$ seen in the SWIG in Figure \ref{fig:front2a} and follows from Assumption \ref{ass:exchAM}\footnote{Since there is no unmeasured common causes of exposure-mediator by Assumption \ref{ass:exchAM}, the only path from $A$ to $Y^{a^\dagger}$ is a backdoor path via $U$.}; line 3 follows by consistency that is implied from recursive substitution of underlying NPSEMs; line 5 follows from Assumption \ref{ass:exchAM}\footnote{If $U$ has a direct arrow to $M$ not through $A$, then this will violate Assumption \ref{ass:exchAM}.}; line 6 follows from Assumptions \ref{ass:noDirect} and \ref{ass:exchMY} and can be seen from the conditional independence of $Y$ and $A$ given $U$ and $M$ as seen in DAG \ref{fig:front1}\footnote{Assumption \ref{ass:noDirect} ensures that there is no direct path from $A$ to $Y$ not mediated by $M$. In addition, since there is no unmeasured common causes of mediator-outcome by Assumption \ref{ass:exchMY}, the only path from $A$ to $Y^{a^\dagger}$ is a backdoor path via $U$ and a frontdoor path via $M$.}; and line 7 follows from the SWIG in Figure \ref{fig:front2b} where it can be seen that $E(Y^m) = \sum_u E(Y\mid U=u, M=m)f(u) = \sum_a E(Y\mid A=a, M=m)f(a)$ as $U$ or $A$ blocks the backdoor back from $M$ to $Y^m$. \allowdisplaybreaks
Alternatively, line 7 holds by algebraically by realizing the following:
\begin{align*}
\sum_a E(Y\mid A=a, M=m)f(a) &= \sum_{a,u} E(Y\mid A=a, M=m, U=u)f(u\mid a,m)f(a)
\\& = \sum_{u} E(Y\mid U=u, M=m)\sum_{a}f(u\mid a)f(a)
\\& = \sum_u E(Y\mid U=u, M=m)f(u).
\end{align*}

\begin{remark}[The front door formula is a weighted average]
\label{sec:frontdoorweight}
Consider a binary treatment $A$ taking values $a^\dagger$ and $a^\circ$. It can be trivially shown that $E(Y^{a^\dagger})$ is a weighted average of $E(Y\mid A=a^\dagger)$ and a separable estimand of treatment on $Y$ denoted by $E(Y^{a_M=a^\dagger, a_Y=a^\circ})$ that can be identified from the observed data that e.g., follow an extended DAG seen in Figure \ref{fig:front5a_appendix}. 
This extended DAG results from splitting the treatment node $A$ into two sub-components, namely $A_M$ and $A_Y$. The bolded arrows from $A$ to $A_M$ and $A_Y$ indicate a deterministic relationship.\footnote{Specifically in the observed data, $A = A_M = A_Y = 1$ with probability 1, and $A = A_M = A_Y = 0$ with probability 1.}
More specifically, the aforementioned two estimands are weighted by the probability of receiving treatment $a^\dagger$ and $a^\circ$ such that $E(Y^{a^\dagger})$ equals 
\begin{align}
P(A=a^\dagger) E(Y\mid A=a^\dagger) 
 + P(A=a^\circ) \underbrace{\sum_{m} E(Y\mid A=a^\circ, M=m) f(m\mid a^\dagger)}_{E(Y^{a_M=a^\dagger, a_Y=a^\circ})},
\end{align}
as stated in the main text (Equation  \eqref{eq:weightedFDF}).
We utilize this decomposition in deriving efficient influence functions for the frontdoor formula.
We note that that both estimands $E(Y\mid A=a^\dagger)$ and $E(Y^{a_M=a^\dagger, a_Y=a^\circ})$ allow for a  direct path from $A$ to $Y$ not mediated through $M$. Moreover, $E(Y\mid A=a^\dagger)$ is identified even if there are unmeasured common causes of $A$ and $Y$.
This decomposition implies that when all individuals in the observed data take treatment $a^\dagger$, then $E(Y^{a^\dagger}) = E(Y\mid A=a^\dagger)$ and when all individuals in the observed data take treatment $a^\circ$, then $E(Y^{a^\dagger}) = E(Y^{a_M=a^\dagger, a_Y=a^\circ})$ (although this would not be identified from observed data unless $f(m\mid a^\dagger)$ is known a priori for all $m$). 
Note that when $E(Y^{a^\dagger}) = E(Y^{a_M=a^\dagger, a_Y=a^\circ})$, $A = A_Y= a^\circ$ for all observations, and thus intervening on $A_Y$ (and creating a mediated path from $A$\contour{black}{$\rightarrow$} $A_Y\rightarrow Y$) is unnecessary.
\end{remark}

\newpage
\section{Identification and estimation of new causally manipulable estimand in absence of $L$}
\label{sec:appendix_nongen}
In this section, we will assume that $L$ is the empty set. 
\begin{theorem}
    The average counterfactual outcome under an intervention on $A_M$ is identified by the frontdoor formula \ref{eq: simple frontdoor formula} under Assumptions \ref{ass:pos}, \ref{ass: delta 1} and \ref{ass:dismiss}, that is,
\begin{align*}
E&(Y^{a_M=a^\dagger} ) = 
 \sum_{m}P(M=m\mid A=a^\dagger)\underbrace{\sum_{a}E(Y \mid A = a,M=m)P(A=a)}_{(\ast\ast)}.
\end{align*}
\end{theorem}

Suppose that the observed data $\mathcal{O}=(A,M,Y)$ follow a law $P$ which is known to belong to $\mathcal{M}=\{P_\theta:\theta\in \Theta\}$, where $\Theta$ is the parameter space. The efficient influence function $\varphi^{\eff}(\mathcal{O})$ for a causal parameter $\Psi\equiv \Psi(\theta)$ in a {non-parametric model $\mathcal{M}_{\text{np}}$ that imposes no restrictions on the law of $\mathcal{O}$ other than positivity} is given by
${d\Psi(\theta_t)}/{dt}\vert_{t=0} = E\{\varphi^{\eff}(\mathcal{O})S(\mathcal{O})\}$, where ${d\Psi(\theta_t)}/{dt}\mid_{t=0}$ is known as the pathwise derivative of the parameter $\Psi$ along any parametric submodel of the observed data distribution indexed by $t$, and $S(\mathcal{O})$ is the score function of the parametric submodel evaluated at $t=0$ \citep{newey1994,Van2000}.

The frontdoor formula can be re-expressed as a weighted average,
\begin{align}
\Psi = P(A=a^\dagger) E(Y\mid A=a^\dagger) 
 + P(A=a^\circ) {\sum_{m} E(Y\mid A=a^\circ, M=m) f(m\mid a^\dagger)},
 \label{eq:weightedFDF}
\end{align}
and thus the efficient influence function can be broken into two components. Using the chain rule, the efficient influence function of $\Psi = \sum_{m} f(m\mid a^\dagger) \sum_a E(Y\mid A=a, M=m)f(a)$  can be derived by finding the efficient influence function of (1) $\psi_1=P(A=a)$, (2) $\psi_2 = E(Y\mid A=a^\dagger) $ and (3) $\psi_3 = {\sum_{m} E(Y\mid A=a^\circ, M=m) f(m\mid a^\dagger)}$. We will use that $\psi_3$ is an established identifying formula for $E(Y^{A_Y=a^\dagger, A_D=a^\circ})$ a term in established identification formula for separable effect, 
which is equal to the same functional of the observed data law $p(o)$ as $E(Y^{a^\circ}\mid A=a^\dagger)$ in the average treatment effect on the treated (ATT) if $a^\circ=0$ and $a^\dagger=1$ \citep{tchetgen2012semiparametric}. 
\sloppy
\begin{theorem}
The efficient influence function $\varphi^{\eff}(\mathcal{O})$ of the frontdoor formula in $\mathcal{M}_{\text{np}}$ is given by
\begin{align*}
\varphi^{\eff}(\mathcal{O})= &\textcolor{red}{\left[I(A=a^\dagger) - P(A=a^\dagger) \right] \psi_2 + P(A=a^\dagger) \frac{I(A=a^\dagger)}{P(A=a^\dagger)}(Y-\psi_2)} +  
\\&\textcolor{blue}{\frac{P(A=a^\circ)}{P(A=a^\dagger)} \left[I(A=a^\circ) \frac{P(A=a^\dagger \mid M)}{P(A=a^\circ\mid M)}\{Y-b_0(M)\} + I(A=a^\dagger)\{b_0(M)-\psi_3\} \right] + }\\&\\&
\textcolor{blue}{[I(A=a^\circ)-P(A=a^\circ)] \psi_3} ,
\end{align*}
where the terms in red are the efficient influence function for $P(A=a^\dagger)\psi_2$, the terms in blue is the efficient influence function for $P(A=a^\circ)\psi_3$, and $b_0(M) = E(Y\mid A=a^\circ, M)$. The efficient influence function can be reduced to the following,
\begin{align}
\varphi^{\eff}(\mathcal{O})=&I(A=a^\dagger)Y + I(A=a^\circ)\psi_3  + \label{eq:eif_rep1}
\\&{\frac{P(A=a^\circ)}{P(A=a^\dagger)} \left[I(A=a^\circ) \frac{P(A=a^\dagger \mid M)}{P(A=a^\circ\mid M)}\{Y-b_0(M)\} + I(A=a^\dagger)\{b_0(M)-\psi_3\} \right]  } - \Psi \nonumber.
\end{align}
which can be re-expressed as:
\begin{align}
\varphi^{\eff}(\mathcal{O})=&I(A=a^\dagger)Y + I(A=a^\circ)\psi_3  +  \label{eq:eif_rep2}
\\&{\left[I(A=a^\circ) \frac{f(M\mid a^\dagger)}{f(M\mid a^\circ)}\{Y-b_0(M)\} + \frac{I(A=a^\dagger)P(A=a^\circ)}{P(A=a^\dagger)}\{b_0(M)-\psi_3\} \right]  } - \Psi, \nonumber
\end{align}
by realizing that ${I(A=a^\circ)P(A=a^\circ)P(A=a^\dagger \mid M)}\{P(A=a^\dagger)P(A=a^\circ\mid M)\}^{-1} = I(A=a^\circ) {f(M\mid a^\dagger)}\{f(M\mid a^\circ)\}^{-1}$.
\end{theorem}
After some algebra, it can be shown that \eqref{eq:eif_rep2} can be written as the form of the efficient influence function for $\Psi$ given by Equation (5) in Theorem 1 in \cite{Fulcher2020} with $C=\emptyset$. Following Theorem 1 in \cite{Fulcher2020}, the semiparametric efficiency bound for $\Psi$ in $\mathcal{M}_{\text{np}}$ is given by var$\left(\varphi^{\eff}\right)$.

\subsubsection{On semiparametric estimators for the frontdoor formula}

Writing the efficient influence function for the frontdoor formula ($\Psi$) given by in Expressions \eqref{eq:eif_rep1} or \eqref{eq:eif_rep2} allows us to construct estimators that guarantee sample-boundedness. 
A weighted iterative conditional expectation (Weighted ICE) estimator that guarantee sample-boundedness is given in the following algorihtm. In what follows, we let $\mathbb{P}_n(X) = n^{-1}\sum_{i=1}^n X_i$ and let $g^{-1}$ denote a known inverse link function\footnote{For instance, if $Y$ is dichotomous, the $g$ is the logit link function}.

\begin{breakablealgorithm}
\renewcommand{\theenumi}{\Alph{enumi}}   
\caption{Algorithm for Weighted ICE (frontdoor formula)}          
\begin{algorithmic} [1]      
\item Non-parametrically compute $P(A=a^\circ)$ and $P(A=a^\dagger)$.
\item Compute the MLEs $\hat{\alpha}$ of ${\alpha}$ from the observed data for the treatment model $P(A=a \mid M;\alpha)$, or compute the MLEs $\hat{\gamma}$ of $\gamma$ from the observed data for the mediator model $P(M=m \mid A;\gamma)$.
\item In the individuals whose $A=a^\circ$, fit a regression model $Q({M};\theta)= g^{-1}\{\theta^T\phi(M)\}$ for $b_0(M)=E(Y \mid M,A=a^\circ)$ where the score function for each observation is weighted by $\hat{W}_1$ where $\hat{W}_1$ equals $$\frac{P(A=a^\circ)P(A=a^\dagger\mid M;\hat{\alpha})}{P(A=a^\dagger)P(A=a^\circ\mid M;\hat{\alpha})}$$ if $\hat{\alpha}$ was estimated in the previous step, or $\hat{W}$ equals $$\frac{f(M\mid A=a^\dagger;\hat{\gamma})}{f(M\mid A=a^\circ;\hat{\gamma})}$$
if $\hat{\gamma}$ was estimated in the previous step.
Moreover, $\phi(M)$ is a known function of $M$.
More specifically, we solve for $\theta$ in the following estimating equations:
\begin{equation*}
   \mathbb{P}_n \left[ I(A=a^\circ)\phi(M)\hat{W}_1\left\{Y - Q({M};\theta)\right\}\right] = 0
\end{equation*}
\item In those whose $A=a^\dagger$, fit an intercept-only model $T(\beta)= g^{-1}(\beta)$ for $\psi_3 = E\{b_0(M)\mid A=a^\dagger\}$, where the score function for each observation is weighted by $$\hat{W}_2 = \frac{P(A=a^\circ)}{P(A=a^\dagger)}.$$ 
More specifically, we solve for $\beta$ in the following estimating equations:
\begin{equation*}
   \mathbb{P}_n \left[ I(A=a^\dagger)\hat{W}_2\left\{Q({M};\hat{\theta}) - T(\beta)\right\}\right] = 0
\end{equation*}
\item Compute $\hat{T} \equiv T(\hat{\beta})$ for all observations.
\item Estimate $\hat{\Psi}_{WICE}=\mathbb{P}_n\{I(A=a^\dagger)Y + I(A=a^\circ)\hat{T}\}$
\end{algorithmic}
\end{breakablealgorithm}
Steps 3 and 4 ensure that the estimates for $\psi_3 = E\{b_0(M)\mid A=a^\dagger\}$ are sample bounded. Step 6 confirms that $\hat{\Psi}_{WICE}$ is a convex combination of $Y$ and estimates for $\psi_3$, both of which are bounded by the range of the outcome $Y$. Thus, $\hat{\Psi}_{WICE}$ will also be sample-bounded. 
For instance if the outcome is binary, then $\hat{\Psi}_{WICE}$ will always be bounded between 0 and 1.
Note that estimators based on Expression \eqref{eq:eif_rep1} are more convenient to construct than estimators based on Expression \eqref{eq:eif_rep2} when (1) $M$ is continuous, and/or (2) there are multiple mediator variables ($M_1$, $M_2$, $M_3$\ldots). 

In Appendix \ref{sec:appendix_robust}, we prove that an estimator based on the efficient influence function given by \eqref{eq:eif_rep1} is doubly robust in the sense that it will be consistent as long as the model for $P(A=a\mid M)$ or the model for $b_0(M)=E(Y\mid A=a^\circ, M)$ is correctly specified, and an estimator based on the efficient influence function given by \eqref{eq:eif_rep2} is doubly robust in the sense that it will be consistent as long as the model for $P(M=m\mid A)$ or the model for $b_0(M)$ is correctly specified.

\newpage
\section{Interventionist identification with the frontdoor formula}
\label{app:sepeff_generalid}

Consider an extended causal DAG, which includes $A$ and also the variable $A_M$, where the bold arrow from $A$ to $A_M$ indicates a
deterministic relationship. 
That is, Figure \ref{fig:front5a_appendix} is the extended DAG of such a split with $V = (U,A,A_M,M,Y)$, and in the observed data, with probability one under $f(v)$, either  $A = A_M = 1$ or $A = A_M = A_Y = 0$.

\allowdisplaybreaks
Here and henceforth we use of ``$(G)$" to indicate that the variables refer to the hypothetical trial where $A_M$ is randomly assigned, as illustrated in \ref{fig:front5b_appendix}. Consider an intervention that sets $A_M(G)$ to $a_M=a^\dagger$.
The average counterfactual outcome under such an intervention is indeed identified as shown in the following:
\begin{align*}
E&(Y^{a_M=a^\dagger} ) = E(Y(G)^{a_M=a^\dagger}) \\&= \sum_{a}E(Y(G)^{a_M=a^\dagger} \mid A_Y(G)= a)P(A_Y(G)=a)\\
& = \sum_{a}E(Y(G)^{a_M=a^\dagger} \mid A_Y(G)= a, A_M(G)=a^\dagger)P(A_Y(G)=a)\\
& = \sum_{a}E(Y(G)\mid A_Y(G)= a, A_M(G)=a^\dagger)P(A_Y(G)=a)\\
& = \sum_{a,m}E(Y(G)\mid A_Y(G)= a, A_M(G)=a^\dagger, M(G)=m)\\& \hspace{2.5em}P(M(G)=m\mid A_Y(G)= a, A_M(G)=a^\dagger) P(A_Y(G)=a)\\ \\
& = \sum_{m}P(M(G)=m\mid A_Y(G)= a^
\dagger, A_M(G)=a^\dagger)\\&\hspace{2.5em}{\sum_{a} E(Y(G) \mid A_Y(G)= a, A_M(G)=a, M(G)=m)P(A_Y(G)=a)}\\ \\
& = \sum_{m}P(M=m\mid A=a^\dagger){\sum_{a}E(Y \mid A = a,M=m)P(A=a)}\\
\end{align*}
Aside from probability laws, we note the following conditions that are used in the proof above: equality 1 follows from consistency that is implied from recursive substitution; equality 3 follows by definition of $G$ and can be seen via d-separation that follows from intervention on $A_M(G)$ in Figure \ref{fig:front5b_appendix}; equality 4 holds by consistency; equality 6 holds by the conditional independence $M(G)\CI A_Y(G)\mid A_M(G)$ and by the conditional independence $Y(G)\CI A_M(G) \mid A_Y(G), M(G)$, both of which follows from d-separation in Figure \ref{fig:front5b_appendix}; 
and equality 7 holds by definition of $G$, consistency and determinism such that the event $\{A_Y(G)=a\}$ is the same as the event $\{A(G)=a\}$ in $G$ and that $\{A_Y=a, A_M=a\}$ is the same as the event $\{A=a\}$ in the observed data.
Of course, since we are not intervening on $A_Y(G)$, we can remove it from the extended DAG as below. This will only require us to define one particular variable $A_M$ that is deterministically equal $A$ in the observed data.


\begin{figure}
    \centering
    \begin{minipage}{1\textwidth}
      \centering
\begin{tikzpicture}
\begin{scope}[every node/.style={draw=none}]
    \node (A) at ( 0, 0 ) {$A$};
    \node (M)   at (3,  0 )  {$M$};
    \node (Y)   at (6 ,0 ) {$Y$};
    \node (U) at (3, 2) {$U$};
\end{scope}
\begin{scope}[>={stealth[black]},
              every node/.style={fill=white,circle},
              every edge/.style={draw=black, thick}]
	\path [->] (A) edge             (M);
	\path [->] (A) edge[bend right] (Y);
	\path [->] (M) edge             (Y);
	\path [->] (U) edge  (A);
	\path [->] (U) edge  (Y);
\end{scope}
\end{tikzpicture}
\end{minipage}
\caption{\label{fig:front4_appendix}DAG of the observed random variables.}
\end{figure}
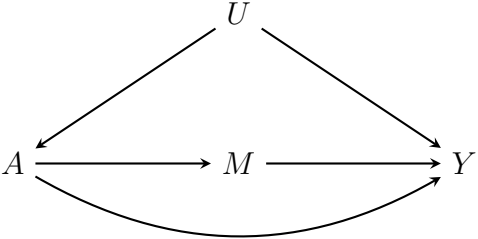

\begin{figure}
    \centering
\begin{minipage}{1\textwidth}
      \centering
\begin{tikzpicture}
\begin{scope}[every node/.style={draw=none}]
    \node (A) at ( 0, -0.75 ) {$A$};
    \node (N) at ( 2, 0 ) {$A_M$};
    \node (O) at ( 2, -1.5 ) {$A_Y$};
    \node (M)   at (4,  0 )  {$M$};
    \node (Y)   at (4 ,-1.5 ) {$Y$};
    \node (U)   at (2 ,-3 ) {$U$};
\end{scope}
\begin{scope}[>={stealth[black]},
              every node/.style={fill=white,circle},
              every edge/.style={draw=black}]
	\path [->, line width=1mm] (A) edge      (N);
	\path [->, line width=1mm] (A) edge      (O);
	\path [->] (N) edge             (M);
	\path [->] (O) edge             (Y);
	\path [->] (M) edge             (Y);
	\path [->] (U) edge [bend left]     (A);
	\path [->] (U) edge [bend right]    (Y);
\end{scope}
\end{tikzpicture}
\subcaption{\label{fig:front5a_appendix}DAG of an expanded graph from Figure \ref{fig:front4_appendix}. Bold arrow denotes deterministic relationship.}
\end{minipage}\\ \vspace{2em}
\begin{minipage}{1\textwidth}
      \centering
\begin{tikzpicture}
\begin{scope}[every node/.style={draw=none}]
    \node (A) at ( 0, -0.75 ) {$A(G)$};
    \node (N) at ( 2, 0 ) {$A_M(G)$};
    \node (O) at ( 2, -1.5 ) {$A_Y(G)$};
    \node (M)   at (5,  0 )  {$M(G)$};
    \node (Y)   at (5 ,-1.5 ) {$Y(G)$};
    \node (U)   at (2 ,-3 ) {$U(G)$};
\end{scope}
\begin{scope}[>={stealth[black]},
              every node/.style={fill=white,circle},
              every edge/.style={draw=black}]
	\path [->, line width=1mm] (A) edge      (O);
	\path [->] (N) edge             (M);
	\path [->] (O) edge             (Y);
	\path [->] (M) edge             (Y);
	\path [->] (U) edge [bend left]     (A);
	\path [->] (U) edge [bend right]    (Y);
\end{scope}
\end{tikzpicture}
\subcaption{\label{fig:front5b_appendix} DAG corresponding to hypothetical trial based on the expanded graph \ref{fig:front5a_appendix}.  Bold arrow denotes deterministic relationship.}
\end{minipage}
\caption{DAG and SWIG of expanded graph.}
\end{figure}
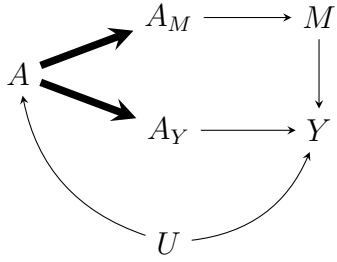
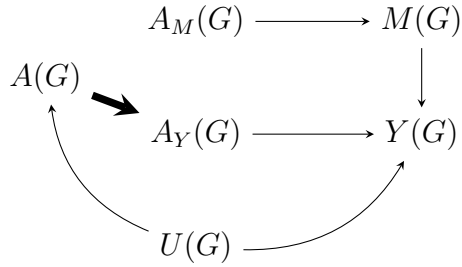

\subsection{Removing $A_Y$ from the extended DAG}
\label{sec:remove_O}
The above results still hold if we remove $A_Y$ from the extended DAG given in Figure \ref{fig:front5a_appendix}. The modified extended DAG with $A_Y$ removed is presented in Figure \ref{fig:front6a}, where again we use of ``$(G)$" to indicate that the variables refer to the hypothetical trial where $A_M$ is randomly assigned. In particular, conditioning sets that include $A(G) = a, A_M(G)=a^\dagger$ refer to the hypothetical trial $G$ in which $A_M$ is randomly assigned\footnote{or imagine a trial where the only arrow into $A_M$ is from $A$ (which may not be deterministic).}. Then, the proof proceeds as follows:
\allowdisplaybreaks
\begin{align*}
E(Y^{a_M=a^\dagger}) &= E(Y(G)^{a_M=a^\dagger})\\&=\sum_{a}E(Y(G)^{a_M=a^\dagger} \mid A(G) = a)P(A(G)=a)\\
& = \sum_{a}E(Y(G)^{a_M=a^\dagger} \mid A(G) = a, A_M(G)=a^\dagger)P(A(G)=a)\\
& = \sum_{a}E(Y(G)\mid A(G) = a, A_M(G)=a^\dagger)P(A(G)=a)\\
& = \sum_{a,m}E(Y(G)\mid A(G) = a, A_M(G)=a^\dagger, M(G)=m)\\&\hspace{2.5em}P(M(G)=m\mid A(G) = a, A_M(G)=a^\dagger) P(A(G)=a)\\ \\
& = \sum_{a,m}P(M(G)=m\mid A(G)=a^\dagger, A_M(G)=a^\dagger)\\&\hspace{2.5em}E(Y(G)\mid A(G) = a, A_M(G)=a, M(G)=m)P(A(G)=a)\\ \\
& = \sum_{m}P(M=m\mid A=a^\dagger)\sum_{a}E(Y\mid A = a, M=m)P(A=a)\\
\end{align*}
\noindent Aside from probability laws, we note the following conditions that are used in the proof above: equality 1 holds by consistency that follows by recursive substitution; equality 3 holds by definition of $G$ and can be seen via d-separation -- $Y(G)^{a_M=a^\dagger} \CI A_M(G) \mid A(G)$ -- that follows by the intervention on $A_M(G)$ in DAG for $G$; equality 4 holds by consistency; equality 6 holds by d-separation following Assumption \eqref{ass:dismiss} where $Y(G)\CI A_M(G) \mid A(G), M(G)$ and $M(G)\CI A(G) \mid A_M(G)$ (since we are assuming for now that $L(G)=\emptyset$); and equality 7 holds by definition of $G$, consistency and determinism in the observed data such that the event $\{A=a, A_M=a\}$ is the same as the event $\{A=a\}$.

\subsection{A causally manipulable interpretation of the Population Intervention Indirect Effect in the general scenario}

As before, consider splitting the treatment node $A$ on a DAG given in Figure \ref{fig:front1} into two sub-components, namely $A_M$ and $A_Y$. Figure \ref{fig:front_gen2a_appendix} is the extended DAG of such a split with $V = (L,U,A,A_M,A_Y,M,Y)$.
This extended DAG is analogous to the extended DAG described in Section \ref{sec:manip_simple} but generalized to include measured common causes ($L$) of $A$, $M$ and $Y$. 
Analogous to Section \ref{sec:remove_O}, we can also remove $A_Y$ in the extended DAG and the results above would still hold to identify the average counterfactual outcome under an intervention on $A_M$ that sets it equal to $a^\dagger$.
Again, consider an intervention on $A_M(G)$ that sets it equal to $a_M=a^\dagger$ as shown in Figure \ref{fig:front_gen2b_appendix}\footnote{we can also imagine a trial where there are non-deterministic arrows from $A$ and $L$ into $A_M$.}.
The average counterfactual outcome under such an intervention is indeed identified in the following:

\allowdisplaybreaks
\begin{align*}
E(Y^{a_M=a^\dagger}) &=  E(Y(G)^{a_M=a^\dagger}) \\&=\sum_{l,a}E(Y(G)^{a_M=a^\dagger} \mid L(G)=l, A(G)=a)P(A(G)=a\mid L(G)=l)P(L(G)=l)\\
& = \sum_{l,a}E(Y(G)^{a_M=a^\dagger} \mid A_M(G)=a^\dagger, L(G)=l, A(G)=a)P(A(G)=a\mid L(G)=l)P(L(G)=l)\\
& = \sum_{l,a}E(Y(G)\mid A_M(G)=a^\dagger, L(G)=l,  A(G)=a){P(A(G)=a\mid L(G)=l)P(L(G)=l)}\\
& = \sum_{m,l,a}E(Y(G)\mid M(G)=m,A_M(G)=a^\dagger, L(G)=l, A(G)=a)\\ & \hspace{2.5em} P(M(G)=m\mid A_M(G)=a^\dagger, A(G)=a, L(G)=l)P(A(G)=a\mid L(G)=l)P(L(G)=l)\\ \\
& = \sum_{m,l}P(M(G)=m\mid A_M(G)=a^\dagger, A(G)=a^\dagger,L(G)=l)P(L(G)=l)\\&\hspace{4em}{\sum_{a}E(Y(G)\mid M(G)=m, L(G)=l, A(G)=a,A_M(G)=a)P(A(G)=a\mid L(G)=l)}\\
& = \sum_{m}P(M=m\mid A=a^\dagger,L=l)P(L=l){\sum_{a}E(Y\mid M=m, L=l, A = a)P(A=a\mid L=l)}\\
\end{align*}

Aside from probability laws, we note the following conditions that are used in the proof above: equality 1 holds by consistency that follows by recursive substitution; equality 3 follows by definition of $G$ and can be seen via d-separation that follows from intervention on $A_M(G)$ in $G$ shown in Figure \ref{fig:front_gen2b_appendix} (which also holds if there are arrows from $L(G)$ to $A_M(G)$ and $A(G)$ to $A_M(G)$); equality 4 holds by consistency; equality 6 holds by the the dismissible component conditions such that conditional independence $M(G)\CI A(G) \mid A_M(G), L(G)$ and $Y(G)\CI A_M(G) \mid A(G), M(G),L(G)$ hold; and equality 7 holds by definition of $G$, consistency and determinism in the observed data such that the event $\{A=a, A_M=a\}$ is the same as the event $\{A=a\}$.

\begin{figure}
    \centering
\begin{minipage}{1\textwidth}
      \centering
\begin{tikzpicture}
\begin{scope}[every node/.style={draw=none}]
    \node (A) at ( 0, -0.75 ) {$A$};
    \node (N) at ( 2, 0 ) {$A_M$};
    \node (O) at ( 2, -1.5 ) {$A_Y$};
    \node (M)   at (4,  0 )  {$M$};
    \node (Y)   at (4 ,-1.5 ) {$Y$};
    \node (U)   at (2 ,-3 ) {$U$};
    \node (L)   at (-2 ,-0.75 ) {$L$};
\end{scope}
\begin{scope}[>={stealth[black]},
              every node/.style={fill=white,circle},
              every edge/.style={draw=black}]
\path [->, line width=1mm] (A) edge      (N);
\path [->, line width=1mm] (A) edge      (O);
\path [->] (N) edge             (M);
\path [->] (O) edge             (Y);
\path [->] (M) edge             (Y);
\path [->] (U) edge [bend left] (A);
\path [->] (U) edge [bend right] (Y);
\path [->] (L) edge             (A);
\path [->] (L) edge [bend left] (M);
\path [->] (L) edge [bend right] (Y);
\end{scope}
\end{tikzpicture}
\subcaption{\label{fig:front_gen2a_appendix} DAG of an extended graph from Figure \ref{fig:front1}.}
\end{minipage}\\ \vspace{2em}
\begin{minipage}{1\textwidth}
      \centering
\begin{tikzpicture}
\begin{scope}[every node/.style={draw=none}]
    \node (A) at ( 0, -0.75 ) {$A(G)$};
    \node (N) at ( 2, 0 ) {$A_M(G)$};
    \node (M)   at (5,  0 )  {$M(G)$};
    \node (Y)   at (5 ,-1.5 ) {$Y(G)$};
    \node (U)   at (2 ,-3 ) {$U(G)$};
    \node (L)   at (-2 ,-0.75 ) {$L(G)$};
\end{scope}
\begin{scope}[>={stealth[black]},
              every node/.style={fill=white,circle},
              every edge/.style={draw=black}]
\path [->] (A) edge      (Y);
\path [->] (N) edge             (M);
\path [->] (M) edge             (Y);
\path [->] (U) edge [bend left] (A);
\path [->] (U) edge [bend right] (Y);
\path [->] (L) edge             (A);
\path [->] (L) edge [bend left] (M);
\path [->] (L) edge [bend right] (Y);
\end{scope}
\end{tikzpicture}
\subcaption{\label{fig:front_gen2b_appendix} SWIG corresponding to an intervention on the extended graph in \ref{fig:front_gen2a_appendix}.}
\end{minipage}
\caption{extended DAG and SWIG.}
\end{figure}

\newpage
\section{Derivation of efficient influence function in Section \ref{sec:estimators}}
\label{sec:EIF_deriv}

\begin{proof}
Note that for binary treatment variables, the generalized frontdoor formula $\Psi = \sum_{m} f(m\mid a^\dagger) \sum_a E(Y\mid A=a, M=m)f(a)$ is equivalent to the following:
$$\Psi = P(A=a^\dagger)E(Y\mid A=a^\dagger) + P(A=a^\circ)\sum_{m,l}E(Y\mid M=m, L=l, A=a^\circ)f(m\mid a^\dagger,l)f(l\mid a^\circ). $$
The efficient influence function in the nonparametric model $\mathcal{M}_{NP}$ is defined as the unique mean zero, finite variance random variable $\varphi^{\eff}(\mathcal{O})$ such that
\[
\frac{d\Psi(\theta_t)}{dt}\Big\vert_{t=0} = E\{\varphi^{\eff}(\mathcal{O})S(\mathcal{O})\}
\]
where $\mathcal{O}=(L,A,M,Y)$, ${d\Psi(\theta_t)}/{dt}\mid_{t=0}$ is known as the pathwise derivative of parameter $\Psi$ along a parametric submodel indexed by $t$, and $S(\mathcal{O})$ is the score function of the parametric submodel evaluated at $t=0$.

The efficient influence function of $\Psi$ can be realized by finding the efficient influence function of (1) $\psi_1=P(A=a)$, (2) $\psi_2 = E(Y\mid A=a^\dagger) $ and (3) $\psi_3 = \sum_{m,l}E(Y\mid M=m, L=l, A=a^\circ)f(m\mid a^\dagger,l)f(l\mid a^\circ)$.  In particular, using differentiation rules, the efficient influence function is given by:
\begin{align*}
\varphi^{\eff}(\mathcal{O}) = \underbrace{[I(A=a^\dagger) - P(A=a^\dagger)]}_{(\ast)}\psi_2 + \psi_2^{\text{eff}}P(A=a^\dagger) + \psi_3^{\text{eff}}P(A=a^\circ) + \underbrace{[I(A=a^\circ) - P(A=a^\circ)]}_{(\ast\ast)}\psi_3
\end{align*}
where expression $(\ast)$ is the efficient influence function of $P(A=a^\dagger)$ and expression $(\ast\ast)$ is the efficient influence function of $P(A=a^\circ)$. Moreover,
\begin{align*}
\frac{d\psi_2(\theta_t)}{dt}\Big\vert_{t=0} &= 
E\{YS({Y\mid A=a^\dagger})\mid A=a^\dagger \}
\\ & = E\left[\left\{Y-E(Y\mid A=a^\dagger)\right\}S({Y\mid A=a^\dagger})\mid A=a^\dagger \right]
\\& = E\left[\frac{I(A=a^\dagger)\left(Y-\psi_2\right)}{P(A=a^\dagger)}S(\mathcal{O})\right] 
\end{align*}
and
\begin{align*}
\frac{d\psi_3(\theta_t)}{dt}\Big\vert_{t=0} = &\underbrace{E\left( E\left[E\{YS(Y\mid A=a^\circ, M,L)\mid A=a^\circ, M,L\}\mid A=a^\dagger,L\right]\mid A=a^\circ\right)}_{\textcircled{\raisebox{-0.05em}{\scalebox{0.85}{A}}}}+ \\
& \underbrace{E\left[E\left\{ E(Y\mid A=a^\circ, M,L)S(M\mid A=a^\dagger,L)\mid A=a^\dagger,L\right\}\mid A=a^\circ\right]}_{\textcircled{\raisebox{-0.05em}{\scalebox{0.85}{B}}}} + \\
& \underbrace{E\left[ E\left\{ E(Y\mid A=a^\circ, M,L)\mid A=a^\dagger,L\right\}S(L\mid A=a^\circ)\mid A=a^\circ\right]}_{\textcircled{\raisebox{-0.05em}{\scalebox{0.85}{C}}}}
\end{align*}
We look at each expression separately. First, we consider  Expression $\textcircled{\raisebox{-0.05em}{\scalebox{0.85}{A}}}$:
\begin{align*}
\textcircled{\raisebox{-0.05em}{\scalebox{0.85}{A}}} &= E\left( E\left[E\{YS(Y\mid A=a^\circ, M,L)\mid A=a^\circ, M,L\}\mid A=a^\dagger,L\right]\mid A=a^\circ\right)\\
& = E\left( E\left[E\left\{Y\frac{I(A=a^\circ)}{P(A=a^\circ\mid L,M)}S(Y\mid A ,M,L)\mid M,L\right\}\mid A=a^\dagger,L\right]\mid A=a^\circ\right)\\
&= E\left( E\left[E\left\{Y\frac{I(A=a^\circ)}{P(A=a^\circ\mid L,M)}S(Y\mid A ,M,L)\mid M,L\right\}\frac{I(A=a^\dagger)}{P(A=a^\dagger\mid L)}\mid L\right] \frac{I(A=a^\circ)}{P(A=a^\circ)}\right)\\
&= E\left( E\left[E\left\{Y\frac{I(A=a^\circ)}{P(A=a^\circ\mid L,M)}S(Y\mid A ,M,L)\mid M,L\right\}\frac{P(A=a^\dagger\mid L,M)}{P(A=a^\dagger\mid L)}\mid L\right] \frac{P(A=a^\circ\mid L)}{P(A=a^\circ)}\right)\\
&= E\left( E\left[E\left\{\Big(Y- b_0(M,L)\Big)\frac{I(A=a^\circ)}{P(A=a^\circ\mid L,M)}S(Y\mid A ,M,L)\mid M,L\right\}\frac{P(A=a^\dagger\mid L,M)}{P(A=a^\dagger\mid L)}\mid L\right] \frac{P(A=a^\circ\mid L)}{P(A=a^\circ)}\right)\\
&= E\left[ \Big\{Y- b_0(M,L)\Big\}\frac{I(A=a^\circ)P(A=a^\dagger\mid L,M)P(A=a^\circ\mid L)}{P(A=a^\circ\mid L,M)P(A=a^\dagger\mid L)P(A=a^\circ)}S(\mathcal{O})\right]\\
\end{align*}
which also equals 
$$ E\left[ \Big\{Y- b_0(M,L)\Big\}\frac{I(A=a^\circ)f(M\mid A=a^\dagger, L)}{P(A=a^\circ)f(M\mid A=a^\circ, L)P(A=a^\dagger\mid L)}S(\mathcal{O})\right]. $$
Next, we consider $\textcircled{\raisebox{-0.05em}{\scalebox{0.85}{B}}}$:
\begin{align*}
\textcircled{\raisebox{-0.05em}{\scalebox{0.85}{B}}} &= E\left[E\left\{ \underbrace{E(Y\mid A=a^\circ, M,L)}_{b_0(M,L)}S(M\mid A=a^\dagger,L)\mid A=a^\dagger,L\right\}\mid A=a^\circ\right]
\\& = E\left[E\left\{ b_0(M,L)\frac{I(A=a^\dagger)}{P(A=a^\dagger\mid L)}S(M\mid A,L)\mid L\right\} \frac{P(A=a^\circ\mid L)}{P(A=a^\circ)}\right]
\\& = E\left[ \Big\{b_0(M,L)-E\Big(b_0(M,L)\mid L,A\Big)\Big\}\frac{I(A=a^\dagger)P(A=a^\circ\mid L)}{P(A=a^\dagger\mid L)P(A=a^\circ)}S(\mathcal{O})\right]
\end{align*}
Finally, we consider $\textcircled{\raisebox{-0.05em}{\scalebox{0.85}{C}}}$:
\begin{align*}
\textcircled{\raisebox{-0.05em}{\scalebox{0.85}{C}}} & = E\left[ \underbrace{E\left\{ E(Y\mid A=a^\circ, M,L)\mid A=a^\dagger,L\right\}}_{h_\dagger(L)}S(L\mid A=a^\circ)\mid A=a^\circ\right]
\\& = E\left\{h_\dagger(L)\frac{I(A=a^\circ)}{P(A=a^\circ)}S(L\mid A) \right\}
\\& = E\left\{\Big(h_\dagger(L)-\psi_3\Big)\frac{I(A=a^\circ)}{P(A=a^\circ)}S(\mathcal{O}) \right\}
\end{align*}
Thus, putting everything together and after some further algebraic simplification, we can see that the efficient influence function is indeed given by:
\begin{align*}
\varphi^{\eff}(\mathcal{O})=&I(A=a^\dagger)Y + I(A=a^\dagger)\psi_3  + 
\\&{\frac{P(A=a^\circ)}{P(A=a^\dagger)} \left[I(A=a^\circ) \frac{P(A=a^\dagger \mid M)}{P(A=a^\circ\mid M)}\{Y-b_0(M)\} + I(A=a^\dagger)\{b_0(M)-\psi_3\} \right]  } - \Psi 
\end{align*}
It is trivial to realize that this Expression of the efficient influence function can also be re-expressed as the following:
\begin{align*}
\varphi^{\eff}(\mathcal{O})=&I(A=a^\dagger)Y + I(A=a^\dagger)\psi_3  +  
\\&{\left[I(A=a^\circ) \frac{f(M\mid a^\dagger)}{f(M\mid a^\circ)}\{Y-b_0(M)\} + \frac{I(A=a^\dagger)P(A=a^\circ)}{P(A=a^\dagger)}\{b_0(M)-\psi_3\} \right]  } - \Psi
\end{align*}
\end{proof}

\newpage
\section{Robustness against model misspecification}
\label{sec:appendix_robust}
\subsection{Non-generalized front door formula}
We show that an estimator based on Equation \eqref{eq:eif_rep1} is doubly robust in the sense that it will be consistent as long as
\begin{enumerate}
\item the model for $P(A=a\mid M)$ is correctly specified, or
\item the model for $E(Y\mid A=a^\circ, M)$ is correctly specified.
\end{enumerate} 
and that an estimator based on Equation \eqref{eq:eif_rep2} is doubly robust in that it will be consistent as long as
\begin{enumerate}
\item the model for $P(M=m\mid A)$ is correctly specified, or
\item the model for $E(Y\mid A=a^\circ, M)$ is correctly specified.
\end{enumerate} 

We consider an estimator based on Equation \eqref{eq:eif_rep1}. Suppose that $\alpha^\ast$, $\theta^\ast$ and $\beta^\ast$ are probability limits of $\alpha$, $\theta$ and $\beta$, respectively. Furthermore, let $b_0^\ast(M) = Q(M; \theta^\ast)$ where as before $Q(M; \theta)$ is a  model for $b_0(M) = E(Y\mid M, A=a^\circ)$, and let $\psi_3^\ast = T(\beta^\ast)$ where $T(\beta)$ is a non-parametric model for $\psi_3=E\{b_0(M)\mid A=a^\dagger\}$.
Under Equation \eqref{eq:eif_rep1} suffices to show that 
\begin{align*}
&E\Bigg(I(A=a^\dagger)Y + I(A=a^\circ)\psi_3^\ast  +\\&
\frac{P(A=a^\circ)}{P(A=a^\dagger)} \left[I(A=a^\circ) \frac{P(A=a^\dagger \mid M; \alpha^\ast)}{P(A=a^\circ\mid M; \alpha^\ast)}\{Y-b_0^\ast(M) \} + I(A=a^\dagger)\{b_0^\ast(M) -\psi_3^\ast\} \right]   - \Psi\Bigg)=0
\end{align*}
under scenario \textbf{(1)} where $\alpha^\ast = \alpha$ and thus  $P(A=a^\dagger \mid M; \alpha^\ast) = P(A=a^\dagger \mid M)$, \textbf{or} under scenario \textbf{(2)} where  $\theta^\ast = \theta$ and thus $b_0^\ast(M) = b_0(M)$ and $\psi_3^\ast = \psi_3$.

\allowdisplaybreaks
\begin{proof}
Suppose first that only the model for $P(A=a\mid M)$ is correctly specified. Then,
\begin{align*}
&E\Bigg(I(A=a^\dagger)Y + I(A=a^\circ)\psi_3^\ast  +\\&
\frac{P(A=a^\circ)}{P(A=a^\dagger)} \left[I(A=a^\circ) \frac{P(A=a^\dagger \mid M; \alpha^\ast)}{P(A=a^\circ\mid M; \alpha^\ast)}\{Y-b_0^\ast(M) \} + I(A=a^\dagger)\{b_0^\ast(M) -\psi_3^\ast\} \right]   - \Psi\Bigg)
\\=& E\Bigg(P(A=a^\dagger\mid M)E(Y\mid A=a^\dagger,M) + P(A=a^\circ\mid M)\psi_3^\ast  +\\&
\frac{P(A=a^\circ)}{P(A=a^\dagger)} \left[\cancel{P(A=a^\circ\mid M)} \frac{P(A=a^\dagger \mid M; \alpha^\ast)}{\cancel{P(A=a^\circ\mid M; \alpha^\ast)}}\{b_0(M)-b_0^\ast(M) \} + P(A=a^\dagger\mid M)\{b_0^\ast(M) -\psi_3^\ast\} \right]   - \Psi\Bigg)
\\=&\sum_m \Big(P(A=a^\dagger\mid M=m)E(Y\mid A=a^\dagger,M=m) + P(A=a^\circ\mid M=m)\psi_3^\ast  +\\&
\frac{P(A=a^\circ)}{P(A=a^\dagger)} \Big[P(A=a^\dagger \mid M=m; \alpha^\ast)b_0(m) - P(A=a^\dagger\mid M=m)\psi_3^\ast\} \Big]\Bigg)P(M=m)   - \Psi
\\=&E(Y\mid A=a^\dagger)P(A=a^\dagger)  + 
\sum_m P(A=a^\circ)P(M=m\mid A=a^\dagger) b_0(M) - \Psi
\\=&0
\end{align*}
Next, suppose that only the model for $E(Y\mid A=a^\circ, M)$ is correctly specified. Then,
\begin{align*}
&E\Bigg(I(A=a^\dagger)Y + I(A=a^\circ)\psi_3^\ast  +\\&
\frac{P(A=a^\circ)}{P(A=a^\dagger)} \left[I(A=a^\circ) \frac{P(A=a^\dagger \mid M; \alpha^\ast)}{P(A=a^\circ\mid M; \alpha^\ast)}\{Y-b_0^\ast(M) \} + I(A=a^\dagger)\{b_0^\ast(M) -\psi_3^\ast\} \right]   - \Psi\Bigg)
\\=& E\Bigg(P(A=a^\dagger\mid M)E(Y\mid A=a^\dagger,M) + P(A=a^\circ\mid M)\psi_3^\ast  +\\&
\frac{P(A=a^\circ)}{P(A=a^\dagger)} \left[P(A=a^\circ\mid M) \frac{P(A=a^\dagger \mid M; \alpha^\ast)}{P(A=a^\circ\mid M; \alpha^\ast)}\{\underbrace{b_0(M)-b_0^\ast(M)}_{=0} \} + P(A=a^\dagger\mid M)\{b_0^\ast(M) -\psi_3^\ast\} \right]   - \Psi\Bigg)
\\=& \sum_m\Bigg(P(A=a^\dagger\mid M=m)E(Y\mid A=a^\dagger,M=m) + P(A=a^\circ\mid M=m)\psi_3^\ast  +\\&
\frac{P(A=a^\circ)}{P(A=a^\dagger)} \left[ P(A=a^\dagger\mid M=m)\{b_0^\ast(M) -\psi_3^\ast\} \right]\Bigg) P(M=m)   - \Psi
\\=&E(Y\mid A=a^\dagger)P(A=a^\dagger)  + 
\sum_m P(A=a^\circ)P(M=m\mid A=a^\dagger) b_0(m) - \Psi
\\=&0
\end{align*}
\end{proof}
The proof of double robustness for estimators based on Equation \eqref{eq:eif_rep2} follows analogously as the proof shown above.

\subsection{Generalized front door formula}
Our proposed estimator based on the efficient influence function given by \eqref{eq:eif_gen1} and \eqref{eq:eif_gen2} is robust against 3 classes of model misspecification scenarios. Specifically, the weighted ICE estimator where a model for $P(A=a\mid M,L)$ is specified will be consistent when at least one of the following holds:
\begin{enumerate}
    \item the models for $P(A=a\mid M,L)$ \textit{and} $P(A=a\mid L)$ are correctly specified, or
    \item the models for $b_0(M,L)$ and $h_\dagger(L)$ are correctly specified, or
    \item the models for $b_0(M,L)$ and $P(A=a\mid L)$ are correctly specified.
\end{enumerate}
The weighted ICE estimator where a model for $P(M=m\mid A,L)$ is specified will be consistent when at least one of the following holds 
\begin{enumerate}
    \item the models for $P(M=m\mid A,L)$ \textit{and} $P(A=a\mid L)$ are correctly specified, or
    \item the models for $b_0(M,L)$ and $h_\dagger(L)$ are correctly specified, or
    \item the models for $b_0(M,L)$ and $P(A=a\mid L)$ are correctly specified.
\end{enumerate}

We will prove robustness for an estimator based on \eqref{eq:eif_gen2} where a model for $P(M=m\mid A,L)$ is specified. Proof of robustness for estimator based on \eqref{eq:eif_gen1} where a model for $P(A=a\mid M,L)$ is specified follows analogously.

Suppose that $\gamma^\ast$, $\kappa^\ast$, $\theta^\ast$, $\eta^\ast$ and $\beta^\ast$ are probability limits of $\gamma$,$\kappa$, $\theta$ $\eta$ and $\beta$, respectively. Furthermore, let $b_0^\ast(M,L) = Q(M,L; \theta^\ast)$ where as before $Q(M,L; \theta)$ is a  model for $b_0(M) = E(Y\mid M, L, A=a^\circ)$, let $h_\dagger^\ast(L) = R(L;\eta^\ast)$ where $R(L;\eta)$ is a model for $h_\dagger(L)$, and let $\psi_3^\ast = T(\beta^\ast)$ where $T(\beta)$ is a non-parametric model for $\psi_3=E\{h_\dagger(L)\mid A=a^\circ\}$.

Under Equation \eqref{eq:eif_gen1} suffices to show that 
\begin{align*}
&E\Bigg(I(A=a^\dagger)Y + I(A=a^\circ)\psi_3^\ast +  \frac{I(A=a^\circ)f(M\mid A=a^\dagger,L;\gamma^\ast)}{f(M\mid A=a^\circ,L;\gamma^\ast)}\{Y-b_0^\ast(M,L)\}  +
\\& \frac{I(A=a^\dagger)P(A=a^\circ\mid L;\kappa^\ast)}{P(A=a^\dagger\mid L;\kappa^\ast)}\{b_0^\ast(M,L) - h_\dagger^\ast(L)\} + I(A=a^\circ)\{h_\dagger^\ast(L) - \psi_3^\ast\} - \Psi
\Bigg)=0
\end{align*}
under scenario \textbf{(1)} where $\gamma^\ast = \gamma$ and $\kappa^\ast=\kappa$ and thus $P(M=m\mid A,L;\gamma^\ast)=P(M=m\mid A,L)$ and  $P(A=a^\dagger \mid L; \kappa^\ast) = P(A=a^\dagger \mid L)$, \textbf{or} under scenario \textbf{(2)} where $\theta^\ast = \theta$ and $\eta^\ast = \eta$ and thus $b_0^\ast(M,L) = b_0(M,L)$, $h_\dagger^\ast(L)=h_\dagger(L)$ and $\psi_3^\ast = \psi_3$, \textbf{or} under scenario \textbf{(3)} where $\theta^\ast = \theta$ and $\kappa^\ast=\kappa$ and thus $b_0^\ast(M,L) = b_0(M,L)$ and $P(A=a^\dagger \mid L; \kappa^\ast) = P(A=a^\dagger \mid L)$.

\begin{proof}
Suppose first that only the models for $P(M=m\mid A,L)$ \textit{and} $P(A=a\mid L)$ are correctly specified. Then,
\begin{align*}
E&\Bigg(I(A=a^\dagger)Y + I(A=a^\circ)\psi_3^\ast +  \frac{I(A=a^\circ)f(M\mid A=a^\dagger,L;\gamma^\ast)}{f(M\mid A=a^\circ,L;\gamma^\ast)}\{Y-b_0^\ast(M,L)\}  +
\\& \frac{I(A=a^\dagger)P(A=a^\circ\mid L;\kappa^\ast)}{P(A=a^\dagger\mid L;\kappa^\ast)}\{b_0^\ast(M,L) - h_\dagger^\ast(L)\} + I(A=a^\circ)\{h_\dagger^\ast(L) - \psi_3^\ast\} - \Psi
\Bigg)
\\=&E\Bigg(\sum_m P(A=a^\dagger\mid L)E(Y\mid A=a^\dagger,M=m,L)f(m\mid a^\dagger,L) + P(A=a^\circ\mid L)\psi_3^\ast + \\& \sum_m\frac{P(A=a^\circ\mid L)f(m\mid A=a^\dagger,L;\gamma^\ast)}{\cancel{f(m\mid A=a^\circ,L;\gamma^\ast)}}\{b_0(m,L)-b_0^\ast(m,L)\}\cancel{f(m\mid a^\circ,L)}  +
\\& \frac{\cancel{P(A=a^\dagger\mid L)}P(A=a^\circ\mid L;\kappa^\ast)}{\cancel{P(A=a^\dagger\mid L;\kappa^\ast)}}\left\{\sum_mb_0^\ast(m,L)f(m\mid a^\dagger,L) - h_\dagger^\ast(L)\right\} + 
\\& P(A=a^\circ\mid L)\{h_\dagger^\ast(L) - \psi_3^\ast\} - \Psi
\Bigg)
\\=&E\Bigg(\sum_m P(A=a^\dagger\mid L)E(Y\mid A=a^\dagger,M=m,L)f(m\mid a^\dagger,L) +   \sum_m{P(A=a^\circ\mid L)f(m\mid A=a^\dagger,L;\gamma^\ast)}b_0(m,L) \Bigg)
\\&=P(A=a^\dagger)E(Y\mid A=a^\dagger) + P(A=a^\circ)\sum_{m,l}b_0(m,l)f(m\mid a^\dagger,l)f(l\mid a^\circ)
\\=&0
\end{align*}
Next, suppose that only the models for $b_0(M,L)$ and $h_\dagger(L)$ are correctly specified. Then,
\begin{align*}
E&\Bigg(I(A=a^\dagger)Y + I(A=a^\circ)\psi_3^\ast +  \frac{I(A=a^\circ)f(M\mid A=a^\dagger,L;\gamma^\ast)}{f(M\mid A=a^\circ,L;\gamma^\ast)}\{Y-b_0^\ast(M,L)\}  +
\\& \frac{I(A=a^\dagger)P(A=a^\circ\mid L;\kappa^\ast)}{P(A=a^\dagger\mid L;\kappa^\ast)}\{b_0^\ast(M,L) - h_\dagger^\ast(L)\} + I(A=a^\circ)\{h_\dagger^\ast(L) - \psi_3^\ast\} - \Psi
\Bigg)
\\=&E\Bigg(\sum_m P(A=a^\dagger\mid L)E(Y\mid A=a^\dagger,M=m,L)f(m\mid a^\dagger,L) + \cancel{P(A=a^\circ\mid L)\psi_3^\ast} + \\& \sum_m\frac{P(A=a^\circ\mid L)f(m\mid A=a^\dagger,L;\gamma^\ast)}{f(m\mid A=a^\circ,L;\gamma^\ast)}\{\underbrace{b_0(m,L)-b_0^\ast(m,L)}_{=0}\}f(m\mid a^\circ,L)  +
\\& \frac{P(A=a^\dagger\mid L)P(A=a^\circ\mid L;\kappa^\ast)}{P(A=a^\dagger\mid L;\kappa^\ast)}\left\{\underbrace{\sum_mb_0^\ast(m,L)f(m\mid a^\dagger,L) - h_\dagger^\ast(L)}_{=0}\right\} + 
\\& P(A=a^\circ\mid L)\{h_\dagger^\ast(L) - \cancel{\psi_3^\ast}\} - \Psi
\Bigg)
\\=&E\Bigg(\sum_m P(A=a^\dagger\mid L)E(Y\mid A=a^\dagger,M=m,L)f(m\mid a^\dagger,L) +   \sum_m{P(A=a^\circ\mid L)f(m\mid A=a^\dagger,L;\gamma^\ast)}b_0(m,L) \Bigg)
\\=&P(A=a^\dagger)E(Y\mid A=a^\dagger) + P(A=a^\circ)\sum_{m,l}b_0(m,l)f(m\mid a^\dagger,l)f(l\mid a^\circ)
\\=&0
\end{align*}

Finally, suppose that only the models for $b_0(M,L)$ and $P(A=a\mid L)$ are correctly specified. Then,
\begin{align*}
E&\Bigg(I(A=a^\dagger)Y + I(A=a^\circ)\psi_3^\ast +  \frac{I(A=a^\circ)f(M\mid A=a^\dagger,L;\gamma^\ast)}{f(M\mid A=a^\circ,L;\gamma^\ast)}\{Y-b_0^\ast(M,L)\}  +
\\& \frac{I(A=a^\dagger)P(A=a^\circ\mid L;\kappa^\ast)}{P(A=a^\dagger\mid L;\kappa^\ast)}\{b_0^\ast(M,L) - h_\dagger^\ast(L)\} + I(A=a^\circ)\{h_\dagger^\ast(L) - \psi_3^\ast\} - \Psi
\Bigg)
\\=&E\Bigg(\sum_m P(A=a^\dagger\mid L)E(Y\mid A=a^\dagger,M=m,L)f(m\mid a^\dagger,L) + \cancel{P(A=a^\circ\mid L)\psi_3^\ast} + \\& \sum_m\frac{P(A=a^\circ\mid L)f(m\mid A=a^\dagger,L;\gamma^\ast)}{f(m\mid A=a^\circ,L;\gamma^\ast)}\{\underbrace{b_0(m,L)-b_0^\ast(m,L)}_{=0}\}f(m\mid a^\circ,L)  +
\\& \frac{\cancel{P(A=a^\dagger\mid L)}P(A=a^\circ\mid L;\kappa^\ast)}{\cancel{P(A=a^\dagger\mid L;\kappa^\ast)}}\left\{\sum_mb_0^\ast(m,L)f(m\mid a^\dagger,L) - h_\dagger^\ast(L)\right\} + 
\\& P(A=a^\circ\mid L)\{h_\dagger^\ast(L) - \cancel{\psi_3^\ast}\} - \Psi
\Bigg)
\\ =&E\Bigg(\sum_m P(A=a^\dagger\mid L)E(Y\mid A=a^\dagger,M=m,L)f(m\mid a^\dagger,L) + 
\\& P(A=a^\circ\mid L;\kappa^\ast)\left\{\sum_mb_0^\ast(m,L)f(m\mid a^\dagger,L) -\cancel{h_\dagger^\ast(L)}\right\} + \cancel{P(A=a^\circ\mid L)h_\dagger^\ast(L)} - \Psi
\Bigg)
\\=&P(A=a^\dagger)E(Y\mid A=a^\dagger) + P(A=a^\circ)\sum_{m,l}b_0(m,l)f(m\mid a^\dagger,l)f(l\mid a^\circ)
\\=&0
\end{align*}

\end{proof}

\newpage

\section{Other relevant estimators}
\label{sec:appendix_estimators}

\subsection{Inverse probability weighted estimator}
We describe one class of inverse probability weighted estimator that was used in the simulation and data analysis (see \citealp{Fulcher2020} for other inverse probability weighted estimators). Specifically, we can solve for $\Psi_{IPW}$ in the following IPW estimator to estimate $\Psi$:
\begin{equation*}
   \mathbb{P}_n \left[\frac{I(A=a^\dagger)}{f(A\mid L;\hat{\kappa})}\left\{\sum_a {E}(Y\mid A=a, M,L;\hat{\theta})f(a\mid L;\hat{\kappa})-\Psi_{IPW}\right\}  \right] = 0
\end{equation*}
where $\mathbb{P}_n(X) = n^{-1}\sum_{i=1}^n X_i$ and ${E}(Y\mid A, M,L;\hat{\theta})$ is an estimate of ${E}(Y\mid A, M,L)$ such that ${E}(Y\mid A=a^\circ, M,L) = b_0(M,L)$.  

\subsection{Iterative conditional expectation estimator}
The ICE estimator that was used in the simulation and data analysis follows from the weighted ICE procedure, whereby we set $\hat{W}=1$ for all regression steps. 

\subsection{Iterative TMLE}
Herein, we describe an iterative TMLE algorithm that is doubly robust in the sense of \cite{Fulcher2020} for binary mediators. 

\begin{breakablealgorithm}
\renewcommand{\theenumi}{\Alph{enumi}}   
\caption{Algorithm for iterative Targeted maximum likelihood (generalized frontdoor formula)} 
\begin{algorithmic} [1]      
\item Non-parametrically compute $P(A=a^\circ)$ and $P(A=a^\dagger)$.
\item Obtain estimates $\hat P(A=a \mid L)$, $\hat P(A=a \mid M,L)$ and $\hat{R}^{(0)}(A,L) \coloneqq \hat P(M=m \mid A, L)$ of $ P(A=a \mid L)$, $P(A=a \mid M,L)$ and $P(M=m \mid A, L)$, respectively, possibly using machine learning methods.
\item In the individuals whose $A=a^\circ$, compute $\hat{Q}^{(0)}({M,L})$ by regressing $Y$ on $(M,L)$. Here, $\hat{Q}^{(0)}({M,L})$ is possibly estimated using machine learning methods, and it denotes an initial estimate for $b_0(M,L)=E(Y \mid M,L,A=a^\circ)$.
\item Define $t=0$. Iteratively update the following until convergence (i.e., until parameters $\delta^{(K)}\approx 0$ and $\nu^{(K)}\approx 0$):
\begin{enumerate}
    \item Update the previous regression. Specifically, in the individuals whose $A=a^\circ$, fit a intercept-only regression model $Q^{(t+1)}({M,L};\delta^{(t+1)})=  g^{-1}[g\{\hat{Q}^{(t)}({M,L})\}+\delta^{(t+1)}]$ where the score function for each observation is weighted by $\hat{W}_1^{(t)} = \frac{\hat f^{(t)}(M\mid A=a^\dagger, L)}{\hat f(M\mid A=a^\circ, L)}$. More specifically, we solve for $\delta$ in the following estimating equations:
    \sloppy
\begin{equation*}
   \mathbb{P}_n \left[ I(A=a^\circ)\hat{W}_1^{(t)} \left\{Y - Q^{(t+1)}({M,L};\delta^{(t+1)})\right\}\right] = 0
\end{equation*}
    \item Define $\hat Q^{(t+1)}_{\text{diff}} =  Q^{(t+1)}({m=1,L};\hat\delta^{(t+1)})-Q^{(t+1)}({m=0,L};\hat\delta^{(t+1)})$. In those whose $A=a^\dagger$, fit a single covariate regression model (with no intercept) for the conditional distribution of $M$ given by $$R^{(t+1)}(a^\dagger, L;\nu^{(t+1)})= g^{-1[}\{g\{\hat{R}^{(t)}(a^\dagger,L)\}+\nu^{(t+1)}\hat Q^{(t+1)}_{\text{diff}}]$$ with observational weights given by $\hat W_2 = \frac{\hat P(A=a^\circ\mid L)}{\hat P(A=a^\dagger\mid L)}$.
More specifically, we solve for $\nu^{(t+1)}$ in the following estimating equations:
\begin{equation*}
   \mathbb{P}_n \left[ I(A=a^\dagger)\hat W_2\hat Q^{(t+1)}_{\text{diff}}\left\{M - R^{(t+1)}(a^\dagger, L;\nu^{(t+1)})\right\}\right] = 0
\end{equation*}
\item Update $\hat Q^{(t+1)}({M,L}) \coloneqq Q^{(t+1)} ({M,L};\hat\delta^{(t+1)})$ and $\hat R^{(t+1)}(a^\dagger, L) \coloneqq R^{(t+1)}(a^\dagger, L;\hat\nu^{(t+1)})$. 
\item t+=1
\end{enumerate}
\item Upon convergence at iteration $K$, define $\hat b_0^{(K)}(m,L) = \hat Q^{(K)}({m,L})$ for $m=0,1$ and $\hat f^{(K)}(M\mid A=a^\dagger, L) = \hat R^{(K)}(a^\dagger, L)$. 
In those whose $A=a^\circ$, fit another regression model $T(\beta)= g^{-1}(\beta)$ for $\psi_3 = E\{h_\dagger(L)\mid A=a^\circ\}$ with just an intercept. More specifically, we solve for $\beta$ in the following estimating equations:
\begin{equation*}
   \mathbb{P}_n \left[ I(A=a^\circ)\left\{\sum_{m=0}^1 \hat b_0^{(K)}(m,L)\hat f^{(K)}(m\mid a^\dagger, L)  - T(\beta)\right\}\right] = 0
\end{equation*}
\item Compute $\hat{T} \coloneqq T(\hat{\beta})$ for all observations.
\item Estimate $\hat{\Psi}_{iTMLE}=\mathbb{P}_n\{I(A=a^\dagger)Y + I(A=a^\circ)\hat{T}\}$
\end{algorithmic}
\end{breakablealgorithm}

\newpage
\section{Extensions to discrete exposure variables with more than two levels}
\label{sec:appendix_discrete}

Extensions to discrete  exposure variables with more than two levels is straightforward. To see this, we can show that our estimand can be written as follows:
$$\Psi = P(A=a^\dagger)E(Y\mid A=a^\dagger) + \sum_{\forall a^\circ \neq a^\dagger}P(A=a^\circ)\sum_{m,l}E(Y\mid M=m, L=l, A=a^\circ)f(m\mid a^\dagger,l)f(l\mid a^\circ). $$

It then follows that in this extension, the efficient influence function for $\Psi$ is given by:
The efficient influence function $\varphi^{\text{eff}}(\mathcal{O})$ for $A_Y=(L,A,M,Y)$ is given by
\begin{align*}
\varphi^{\text{eff}}(\mathcal{O}) &= I(A=a^\dagger)Y + \sum_{\forall a^\circ \neq a^\dagger}I(A=a^\circ)\psi_3 + 
\\& \sum_{\forall a^\circ \neq a^\dagger}\Bigg[\frac{I(A=a^\circ)P(A=a^\dagger\mid M,L)P(A=a^\circ\mid L)}{P(A=a^\circ\mid M,L)P(A=a^\dagger\mid L)}\{Y-b_0(M,L)\}  +\nonumber
\\& \hspace{3em}\frac{I(A=a^\dagger)P(A=a^\circ\mid L)}{P(A=a^\dagger\mid L)}\{b_0(M,L) - h_\dagger(L)\} + \nonumber
\\& \hspace{3em}I(A=a^\circ)\{h_\dagger(L) - \psi_3\}\Bigg] - \Psi,
\nonumber
\end{align*}
which can also be re-expressed as
\begin{align*}
\varphi^{\text{eff}}(\mathcal{O}) &= I(A=a^\dagger)Y + \sum_{\forall a^\circ \neq a^\dagger}I(A=a^\circ)\psi_3 + 
\\& \sum_{\forall a^\circ \neq a^\dagger}\Bigg[\frac{I(A=a^\circ)f(M\mid A=a^\dagger,L)}{f(M\mid A=a^\circ,L)}\{Y-b_0(M,L)\}  +\nonumber
\\& \hspace{3em}\frac{I(A=a^\dagger)P(A=a^\circ\mid L)}{P(A=a^\dagger\mid L)}\{b_0(M,L) - h_\dagger(L)\} + \nonumber
\\& \hspace{3em}I(A=a^\circ)\{h_\dagger(L) - \psi_3\}\Bigg] - \Psi.
\nonumber
\end{align*}

The weighted estimator will still be sample-bounded, but will need to be slightly modified in the following way:
\begin{breakablealgorithm}
\renewcommand{\theenumi}{\Alph{enumi}}   
\caption{Algorithm for Weighted ICE (generalized frontdoor formula for discrete exposure with more than two levels)}          
\begin{algorithmic} [1]      
\item Non-parametrically compute $P(A=a)$ for all values of $a\in \mathcal{A}$.
\item Compute the MLEs $\hat{\kappa}$ of ${\kappa}$ from the observed data for the treatment model $P(A=a\mid L;\kappa)$. In addition, compute the MLEs $\hat{\alpha}$ of ${\alpha}$ from the observed data for the treatment model $P(A=a \mid M,L;\alpha)$, or compute the MLEs $\hat{\gamma}$ of $\gamma$ from the observed data for the mediator model $P(M=m \mid A, L;\gamma)$
\item For all levels of $a^\circ\in \mathcal{A}$ that is not equal to $a^\dagger$, do the following:
\begin{enumerate}
\item In the individuals whose $A=a^\circ$, fit a regression model $Q_{a^\circ}({M,L};\theta_{a^\circ})=  g^{-1}\{\theta_{a^\circ}^T\phi_{a^\circ}(M,L)\}$ for $b_{0,a^\circ}(M,L)=E(Y \mid M,L,A=a^\circ)$ where the score function for each observation is weighted by $\hat{W}_{1,a^\circ}$ where $\hat{W}_{1,a^\circ}1$ equals $$\frac{P(A=a^\circ\mid L;\hat{\kappa})P(A=a^\dagger\mid M,L;\hat{\alpha})}{P(A=a^\dagger\mid L;\hat{\kappa})P(A=a^\circ\mid M,L;\hat{\alpha})}$$ if $\hat{\alpha}$ was estimated in the previous step, or $\hat{W}_{1,a^\circ}$ equals $$\frac{f(M\mid A=a^\dagger, L;\hat{\gamma})}{f(M\mid A=a^\circ, L;\hat{\gamma})}$$
if $\hat{\gamma}$ was estimated in the previous step.
Moreover, $\phi_{a^\circ}(M,L)$ is a known function of $M$ and $L$.
More specifically, we solve for $\theta$ in the following estimating equations:
\begin{equation*}
   \mathbb{P}_n \left[ I(A=a^\circ)\phi_{a^\circ}(M,L)\hat{W}_{1,a^\circ} \left\{Y - Q_{a^\circ}({M,L};\theta_{a^\circ})\right\}\right] = 0
\end{equation*}
\item In those whose $A=a^\dagger$, fit a regression model $R_{a^\circ}(L;\eta_{a^\circ})= g^{-1}\{\eta_{a^\circ}^T\Gamma_{a^\circ}(L)\}$ for $h_\dagger(L)=E(b_{a^\circ}(M,L) \mid L,A=a^\dagger)$ where the score function for each observation is weighted by $$\hat{W}_{2,a^\circ} = \frac{P(A=a^\circ\mid L;\hat{\kappa})}{P(A=a^\dagger\mid L; \hat{\kappa})}.$$
Here, $\gamma_{a^\circ}(L)$ is a known function of $L$.
More specifically, we solve for $\eta$ in the following estimating equations:
\begin{equation*}
   \mathbb{P}_n \left[ I(A=a^\dagger)\Gamma_{a^\circ}(L)\hat{W}_{2,a^\circ}\left\{Q_{a^\circ}({M,L};\hat{\theta}_{a^\circ}) - R_{a^\circ}(L;\eta_{a^\circ})\right\}\right] = 0
\end{equation*}
\item In those whose $A=a^\circ$, fit another regression model $T(\beta_{a^\circ})=g^{-1}(\beta_{a^\circ})$ for $\psi_3 = E\{h_{\dagger,{a^\circ}}(L)\mid A=a^\circ\}$ with just an intercept. More specifically, we solve for $\beta_{a^\circ}$ in the following estimating equations:
\begin{equation*}
   \mathbb{P}_n \left[ I(A=a^\circ)\left\{R_{a^\circ}({L};\hat{\eta}_{a^\circ}) - T(\beta_{a^\circ})\right\}\right] = 0
\end{equation*}
\item Compute $\hat{T}_{a^\circ} \equiv T(\hat{\beta}_{a^\circ})$ for all observations.
\end{enumerate}
\item Estimate 
$\hat{\Psi}_{WICE}=\mathbb{P}_n\left\{I(A=a^\dagger)Y + \sum_{\forall a^\circ \neq a^\dagger}I(A=a^\circ)\hat{T}_{a^\circ}\right\}$
\end{algorithmic}
\end{breakablealgorithm}

\newpage
\section{Additional simulation study}
\label{sec:extrasim}

The data-generating mechanism for our second simulation study and model specifications are provided in Table \ref{table:DGM_extra}. We consider four scenarios to illustrate the robustness of our proposed estimator to model misspecification. We consider four model specification scenarios: (1) all models are correctly specified, (2) only the models for $b_0(M,L)$ and $h_\dagger(L)$ are correctly specified,  (3) only the models for $b_0(M,L)$ and $P(A=a\mid L)$ are correctly specified, and (4) only the models for $P(M=m\mid A,L)$ \textit{and} $P(A=a\mid L)$ are correctly specified. The correct mediator model in the specification scenarios is the one used in the data generation process, and the exposure and outcome models are approximately correctly specified by including pairwise interactions between all the variables to ensure flexibility.

Table \ref{table:simresult_extra} shows the results from the simulation study. As expected by our theoretical derivations, when all of the working models are correctly specified, all of the estimators are nearly unbiased. The AIPW estimator and our proposed weighted ICE estimator are also nearly unbiased in the three model misspecification settings whereas the IPW and ICE estimators are not all unbiased.

\begin{table}[h!]
\centering
\begin{tabular}{||p{3cm} p{12cm} ||} 
 \hline \hline
 \multicolumn{2}{||c||}{Data generating mechanism} \\ [0.5ex] 
 \hline
  $U\sim$ & $\text{Ber}\{0.3\}$  \\ 
  $L_1\sim$ & $\text{Ber}\{0.6\}$  \\
  $L_2\sim$ & $\text{Ber}\{\expit(1+4L_1)\}$  \\ 
  $A\sim$ & $\text{Ber}\{\expit(-1+L_1+2L_2+5L_1L_2+U)\}$  \\ 
  $M\sim$ & $\text{Ber}\{\expit(1+A-3L_1+2L_2-5L_1L_2)\}$  \\ 
  $Y\sim$ & $\text{Ber}\{\expit(-2.25+2A-5M-2AM+2L_1-2L_2-5L_1L_2+U)\}$  \\ 
   \hline
 \multicolumn{2}{||c||}{Model misspecification} \\ [0.5ex]
  \hline
Scenario 2 & $P(M=m\mid L,A; \gamma) = \expit(\gamma_0 + \gamma_1 A + \gamma_2 L_2)$\\
 & $P(A=a\mid L; \kappa) = \expit(\kappa_0 + \kappa_1 L_2)$\\ [1ex]
Scenario 3 & $P(M=m\mid L,A; \gamma) = \expit(\gamma_0 + \gamma_1 A + \gamma_2 L_2)$\\ 
 & $R(L;\theta) = \expit(\eta_0 + \eta_1 L_2)$\\ [1ex]
Scenario 4 & $Q(M,L;\theta) = \expit(\theta_0 + \theta_1 M + \theta_2 L_1 + \theta_3 L_2)$\\
 & $R(L;\theta) = \expit(\eta_0 + \eta_1 L_2 (1-L_1))$\\
 \hline\hline
\end{tabular}
\caption{Data generating mechanism and model misspecifications for scenarios in simulation study.}
\label{table:DGM_extra}
\end{table}

\begin{table}
\centering
\small
\begin{tabular}{ |l | c c c | c c c| c c c | c c c|  }
\hline
& \multicolumn{3}{|c|}{Scenario 1} & \multicolumn{3}{|c|}{Scenario 2}  & \multicolumn{3}{|c|}{Scenario 3}  & \multicolumn{3}{|c|}{Scenario 4} \\
\hline
& Bias  & SE & Bias$_{s}$ & Bias  & SE & Bias$_{s}$ & Bias  & SE & Bias$_{s}$ & Bias  & SE & Bias$_{s}$ \\
\hline
\multicolumn{13}{|c|}{$n=100$}  \\
\hline
IPW	&0.05	&1.16	&4.24	&0.80	&2.13	&\textit{37.41}	&0.05	&1.16	&4.24	&3.40	&1.64	&\textbf{208.00}\\
ICE	&0.05	&1.10	&4.70	&0.05	&1.10	&4.70	&0.64	&2.02	&\textit{31.51}	&0.82	&2.45	&\textit{33.33}\\
AIPW	&0.11	&1.20	&9.56	&-	&-	&-	&0.12	&1.23	&9.64	&0.11	&1.22	&9.16\\
WICE	&0.06	&1.11	&5.06	&0.05	&1.10	&4.76	&0.00	&1.08	&0.00	&0.01	&1.13	&1.18\\ \hline
\multicolumn{13}{|c|}{$n=250$}  \\
\hline
IPW	&-0.03	&0.77	&-4.24	&0.92	&1.98	&\textbf{46.55}	&-0.03	&0.77	&-4.24	&3.43	&1.01	&\textbf{338.39}\\
ICE	&-0.01	&0.70	&-1.28	&-0.01	&0.70	&-1.28	&0.81	&1.93	&\textbf{42.10}	&1.09	&2.61	&\textbf{41.87}\\
AIPW	&0.00	&0.70	&0.31	&-	&-	&-	&0.00	&0.73	&0.64	&0.00	&0.70	&-0.11\\
WICE	&-0.01	&0.70	&-1.32	&-0.01	&0.70	&-1.07	&-0.04	&0.70	&-6.09	&-0.04	&0.69	&-6.40\\
\hline
\multicolumn{13}{|c|}{$n=500$}  \\
\hline
IPW	&0.01	&0.93	&1.17	&1.16	&2.19	&\textbf{52.69}	&0.01	&0.93	&1.17	&3.45	&0.73	&\textbf{471.59}\\
ICE	&-0.01	&0.48	&-1.46	&-0.01	&0.48	&-1.46	&0.95	&1.91	&\textbf{49.67}	&1.21	&2.65	&\textbf{45.61}\\
AIPW	&0.00	&0.49	&0.77	&-	&-	&-	&0.02	&0.57	&4.17	&0.00	&0.48	&-0.66\\
WICE	&-0.01	&0.48	&-1.21	&-0.01	&0.48	&-1.20	&-0.02	&0.49	&-4.29	&-0.02	&0.49	&-4.80\\
\hline
\end{tabular}
\caption{Simulation results: Bias, standard error (SE), and standardized bias (Bias$_s$) are multiplied by 100.} 
\label{table:simresult_extra}
\end{table}

\newpage
\section{Asymptotic properties}
\label{sec:asymp}
In observational studies, model misspecification in the estimation of nuisance functions can induce biased estimates of the ACE.        
In recent years, there has been an explosion in developing flexible data-adaptive methods (e.g. kernel smoothing, generalized additive models, ensemble learners, random forest) combined with doubly robust estimators that can reduce the risk of model misspecification and provide valid causal inference.
These machine learning techniques offer more protection against model misspecficiation than the parametric models.

From first order expansion of a singly-robust plug-in estimator (IPW and ICE estimators), it can be shown that we require the nuisance parameter estimators to  converge to the truth at rate $n^{-1/2}$.  However, this is not possible for non-parametric conditional mean functions as this rate is not attainable for these types of functions.
However when doubly robust estimators are used with data-adaptive methods this issue largely disappears are doubly robust estimators enjoy the small bias property \citep{newey2004}.

In this section we will examine the Remainder or Bias term from the following decomposition. 
For notational brevity, we suppress $\mathcal{O}$ in the equations below. 
For generality, suppose that $\Psi(\hat P)$ is an estimator that solves the estimating equations based on the efficient influence function. We have that
\begin{align*}
\sqrt{n} (\Psi(\hat P) - \Psi(P)) &= \sqrt{n}\left[\mathbb{P}_n(\varphi^{eff}(\hat P)) - P(\varphi^{eff}(\hat P)) \right]  + \sqrt{n}\left[\Psi(\hat P) + P(\varphi^{eff}(\hat P)) -\Psi(P) \right]
\\& = \mathbb{G}_n({\varphi}^{eff}({P})) + \mathbb{G}_n[{\varphi}^{eff}(\hat{P}) - {\varphi}^{eff}({P})] + 
\\& ~~\sqrt{n}\left[\Psi(\hat P) + P({\varphi}^{eff}(\hat{P})) - \Psi(P)\right]
\\& = \underbrace{\mathbb{G}_n({\varphi}({P}))}_{T_1} + \underbrace{\mathbb{G}_n[{\varphi}(\hat{P}) - {\varphi}({P})]}_{T_2} + 
\\& ~~\sqrt{n}\left[\underbrace{\Psi(\hat P) + P({\varphi}^{eff}(\hat{P})) - \Psi(P)}_{R}\right]
\end{align*}
where $\mathbb{G}_n[X] = \sqrt{n}(\mathbb{P}_n - P)(X)$ for any $X$ and we define ${\varphi}(\mathcal{O};\tilde P) = {\varphi}^{eff}(\mathcal{O};\tilde P)+\Psi(O;\tilde P)$ for any $\tilde P$.
The first term given by $T_1$ is a centered sample average which converges to a mean zero Normal distribution by the central limit theorem. The second term is known as an empirical process term, which can be shown to be $o_p(1)$ if we assume that nuisance functions and their corresponding estimators are not too complex and belong to Donsker class. Alternatively, one can use sample splitting and cross fitting to overcome issues with overfitting \citep{Chernozhukov2018}. 

Formally, we assume the following conditions for the first two terms:

\newcommand\norm[1]{\lVert#1\rVert}
\newcommand\normx[1]{\Vert#1\Vert}

\begin{enumerate}[label=C\arabic*.]
\itemsep1em 
    \item $E\left[\varphi^{\text{eff}}(\mathcal{O})^2 \right]<\infty$
    \item $\varphi^{\text{eff}}(\mathcal{O})$  and $\hat\varphi^{\text{eff}}(\mathcal{O})$ belong to a Donsker family.
    \item $\norm{\hat\varphi^{\text{eff}}(\mathcal{O}) - \varphi^{\text{eff}}(\mathcal{O})}_2^2 \overset{p}\longrightarrow 0$
\end{enumerate}
where $\hat\varphi^{\text{eff}}(\mathcal{O})$ denotes an estimator of $\varphi^{\text{eff}}(\mathcal{O})$ where all nuisance functions estimators are exactly the same as those in the TMLE estimator. It is not hard to show that $\mathbb{P}_n \{\hat\varphi^{\text{eff}}(\mathcal{O})\}=0$ by construction.

The last term is known as the remainder or bias term. We will need to show that $R = o_p(1)$ under some conditions about the convergence rates of the nuisance functions. 

\small
\begin{align*}
&\Psi(\hat P) + P({\varphi}^{eff}(\hat{P})) - \Psi(P) =
\\& E_P\left[\underbrace{\textcolor{blue}{\frac{I(A=a^\circ)\hat f(M\mid a^\dagger, L)}{\hat f(M\mid a^\circ, L)}(Y-\hat b_0(M,L))}}_{(A)}+\underbrace{\textcolor{red}{\frac{I(A=a^\dagger)\hat f(a^\circ\mid  L)}{\hat f(a^\dagger\mid L)}(\hat b_0(M,L)-\hat h_{\dagger}(L))+I(A=a^\circ)\hat h_{\dagger}(L)-P(A=a^\circ)\psi_3}}_{(B)} \right]
\end{align*}\normalsize

We examine the terms in blue (henceforth denoted as (A)) and term in red (henceforth denoted as (B)) in detail. Starting with the term in (B):

\small
\begin{align*}
(B) &= E_P\left[I(A=a^\circ)(\hat h_{\dagger}(L) - h_{\dagger}(L)) + \frac{I(A=a^\dagger)\hat f(a^\circ\mid  L)}{\hat f(a^\dagger\mid L)}(\hat b_0(M,L)-\hat h_{\dagger}(L)) \right]
\\& = E_P\left[I(A=a^\circ)(\hat h_{\dagger}(L) - h_{\dagger}(L)) + \frac{I(A=a^\dagger)\hat f(a^\circ\mid  L)}{\hat f(a^\dagger\mid L)}\left\{E_P(\hat b_0(M,L)\mid A=a^\dagger,L)-\hat h_{\dagger}(L)\right\} \right]
\\& = E_P\Bigg[f(a^\circ\mid L))(\hat h_{\dagger}(L) - h_{\dagger}(L)) + \frac{f(a^\dagger\mid L)\hat f(a^\circ\mid  L)}{\hat f(a^\dagger\mid L)}\left\{E_P(\hat b_0(M,L)\mid A=a^\dagger,L)-\hat h_{\dagger}(L)\right\} +
\\&\hspace{3em} f(a^\circ\mid L)\left\{E_P(\hat b_0(M,L)\mid A=a^\dagger,L) - E_P(\hat b_0(M,L)\mid A=a^\dagger,L)  \right\}\Bigg]
\\& = E_P\left[\left\{E_P(\hat b_0(M,L)\mid A=a^\dagger,L)-\hat h_{\dagger}(L)\right\}\left\{\frac{f(a^\dagger\mid L)\hat f(a^\circ\mid  L)}{\hat f(a^\dagger\mid L)} - f(a^\circ\mid L)\right\} \right] +
\\&\hspace{3em} E_P\left\{E_P\Big(\hat b_0(M,L)-b_0(M,L)\mid A=a^\dagger,L\Big)I(A=a^\circ)  \right\}
\\& = E_p\left[\left\{E_P(\hat b_0(M,L)\mid A=a^\dagger,L)-\hat h_{\dagger}(L)+h_{\dagger}(L) - h_{\dagger}(L)\right\} \left\{\hat f(a^\circ\mid L) - f(a^\circ\mid L)\right\}\frac{1}{\hat f(a^\dagger\mid L)} \right]
\\&\hspace{3em} E_P\left\{E_P\Big(\hat b_0(M,L)-b_0(M,L)\mid A=a^\dagger,L\Big)I(A=a^\circ)  \right\}
\\& = E_P\left[\{h_{\dagger}(L) - \hat h_{\dagger}(L) \} \left\{\hat f(a^\circ\mid L) - f(a^\circ\mid L)\right\}\frac{1}{\hat f(a^\dagger\mid L)}  \right] + \\&\hspace{3em} E_P\left[E_P(\hat b_0(M,L)-b_0(M,L)\mid A=a^\dagger,L)\left\{\hat f(a^\circ\mid L) - f(a^\circ\mid L)\right\}\frac{1}{\hat f(a^\dagger\mid L)} \right]+
\\&\hspace{3em} \underbrace{\textcolor{purple}{E_P\left\{E_P\Big(\hat b_0(M,L)-b_0(M,L)\mid A=a^\dagger,L\Big)I(A=a^\circ)  \right\}}}_{(B.2)}
\end{align*}\normalsize

We keep in mind the term in purple (term (B.2)) as we expand upon term (A):
\small
\begin{align*}
(A) &= E_P\left[\frac{I(A=a^\circ)\hat f(M\mid a^\dagger, L)}{\hat f(M\mid a^\circ, L)}(Y-\hat b_0(M,L))\right]
\\& = E_P\left[\frac{I(A=a^\circ)\hat f(M\mid a^\dagger, L)}{\hat f(M\mid a^\circ, L)}\Big(b_0(M,L)-\hat b_0(M,L)\Big)\right]
\\&  = E_P\left(I(A=a^\circ)E_P\left[\frac{\hat f(M\mid a^\dagger, L)f(M\mid a^\circ, L)}{\hat f(M\mid a^\circ, L) f(M\mid a^\dagger, L)} \left\{b_0(M,L) - \hat{b}_0(M,L)\right\}\mid A=a^\dagger, L \right] \right)
\end{align*}

\normalsize
Now, adding (A) and (B.2) (term in purple) we get the following:
\small
\begin{align*}
&E_P\left(I(A=a^\circ)E_P\left[\left\{\frac{\hat f(M\mid a^\dagger, L)f(M\mid a^\circ, L)}{\hat f(M\mid a^\circ, L) f(M\mid a^\dagger, L)}-1\right\} \left\{b_0(M,L) - \hat{b}_0(M,L)\right\}\mid A=a^\dagger, L\right] \right)
\\& = E_P\left(I(A=a^\circ)E_P\left[\left\{\frac{f(M\mid a^\circ, L)}{f(M\mid a^\dagger, L)}\frac{1}{\hat f(M\mid a^\circ, L)}\right\}\left\{\hat f(M\mid a^\dagger, L) - f(M\mid a^\dagger, L)\right\} \left\{b_0(M,L) - \hat{b}_0(M,L)\right\}\mid A=a^\dagger, L\right] \right) + 
\\& \hspace{1em}E_P\left(I(A=a^\circ)E_P\left[\left\{\frac{1}{\hat f(M\mid a^\circ, L)}\right\}\left\{ f(M\mid a^\circ, L) - \hat f(M\mid a^\circ, L)\right\} \left\{b_0(M,L) - \hat{b}_0(M,L)\right\}\mid A=a^\dagger, L \right] \right)
\end{align*}

\normalsize
Thus, together we have (A)+(B) equals:

\small
\begin{align*}
&E_P\left[\{h_{\dagger}(L) - \hat h_{\dagger}(L) \} \left\{\hat f(a^\circ\mid L) - f(a^\circ\mid L)\right\}\frac{1}{\hat f(a^\dagger\mid L)}  \right] + 
\\& E_P\left[E_P(\hat b_0(M,L)-b_0(M,L)\mid A=a^\dagger,L)\left\{\hat f(a^\circ\mid L) - f(a^\circ\mid L)\right\}\frac{1}{\hat f(a^\dagger\mid L)} \right]+
\\&
E_P\left(I(A=a^\circ)E_P\left[\left\{\frac{f(M\mid a^\circ, L)}{f(M\mid a^\dagger, L)}\frac{1}{\hat f(M\mid a^\circ, L)}\right\}\left\{\hat f(M\mid a^\dagger, L) - f(M\mid a^\dagger, L)\right\} \left\{b_0(M,L) - \hat{b}_0(M,L)\right\}\mid A=a^\dagger, L \right] \right) + 
\\& E_P\left(I(A=a^\circ)E_P\left[\left\{\frac{1}{\hat f(M\mid a^\circ, L)}\right\}\left\{ f(M\mid a^\circ, L) - \hat f(M\mid a^\circ, L)\right\} \left\{b_0(M,L) - \hat{b}_0(M,L)\right\}\mid A=a^\dagger, L\right] \right)\end{align*}
\normalsize

By an application of Cauchy-Schwartz, we can show that as long as:
\begin{enumerate}
\item $\norm{\hat h_\dagger(L)-h_\dagger(L)}~ \norm{\hat f(a^\circ\mid L) - f(a^\circ\mid L)} = O_p(n^{-\nu})$, and
\item $\norm{\hat b_0(M,L)-b_0(M,L)}~ \norm{\hat f(a^\circ\mid L) - f(a^\circ\mid L)} = O_p(n^{-\nu})$, and
\item $\norm{\hat b_0(M,L)-b_0(M,L)}~ \norm{\hat f(M\mid a, L) - f(M\mid a, L)} = O_p(n^{-\nu}),~~\forall a$
\end{enumerate}
for $\nu>1/2$ and where $\norm{f(x)} = \left\{\int |f(x)|^2dP(x)\right\}^{1/2}$, i.e. the $L_2(P)$ norm. Then, $\sqrt{n}\Big\{\Psi(\hat P) + P({\varphi}^{eff}(\hat{P})) - \Psi(P)\Big\} = o_p(1)$.
This can be accomplished, for example, if the nuisance functions are each consistently estimated at a rate of $n^{-1/4}$ or faster. 

Note that $h_\dagger(L) = \sum_m b_0(m,L)f(m\mid a^\dagger,L)$. In our estimators, we propose estimating $h_\dagger(L)$ by regressing $b_0(M,L)$ on $L$ in those whose $A=a^\dagger$ to ensure sample-boundedness. However, if we estimate is $h_\dagger(L)$ by calculating $\sum_m \hat b_0(m,L)\hat f(m\mid a^\dagger,L)$ explicitly (as in \citealp{Fulcher2020}). Under the AIPW estimator of \cite{Fulcher2020} (and the iterative TMLE), the remainder term reduces to the following:
\begin{align*}
& E_P\left[\sum_m \hat b_0(m,L)\left\{f(m\mid A=a^\dagger,L)-\hat f(m\mid A=a^\dagger,L)\right\}\left\{\hat f(a^\circ\mid L) - f(a^\circ\mid L)\right\}\frac{1}{\hat f(a^\dagger\mid L)} \right]+
\\&
E_P\left(I(A=a^\circ)E_P\left[\left\{\frac{f(M\mid a^\circ, L)}{f(M\mid a^\dagger, L)}\frac{1}{\hat f(M\mid a^\circ, L)}\right\}\left\{\hat f(M\mid a^\dagger, L) - f(M\mid a^\dagger, L)\right\} \left\{b_0(M,L) - \hat{b}_0(M,L)\right\}\mid A=a^\dagger, L \right] \right) + 
\\& E_P\left(I(A=a^\circ)E_P\left[\left\{\frac{1}{\hat f(M\mid a^\circ, L)}\right\}\left\{ f(M\mid a^\circ, L) - \hat f(M\mid a^\circ, L)\right\} \left\{b_0(M,L) - \hat{b}_0(M,L)\right\}\mid A=a^\dagger, L\right] \right).\end{align*}

Then we can show that if the model for $f(M\mid A,L)$ is correctly specified, then the remainder term reduces to the following asymptotically\footnote{Note that this is really only possible if $M$ was discrete, otherwise we need to resort to using numerical integration.}:
\small
\begin{align*}
&E_P\left(I(A=a^\circ)E_P\left[\left\{\frac{f(M\mid a^\circ, L)}{f(M\mid a^\dagger, L)}\frac{1}{f^{\ast}(M\mid a^\circ, L)}\right\}\left\{ f^{\ast}(M\mid a^\dagger, L) - f(M\mid a^\dagger, L)\right\} \left\{b_0(M,L) - {b}^{\ast}_0(M,L)\mid A=a^\dagger, L\right\}\right] \right) + 
\\& E_P\left(I(A=a^\circ)E_P\left[\left\{\frac{1}{ f^\ast(M\mid a^\circ, L)}\right\}\left\{ f(M\mid a^\circ, L) -  f^{\ast}(M\mid a^\circ, L)\right\} \left\{b_0(M,L) - {b}^{\ast}_0(M,L)\mid A=a^\dagger, L\right\}\right] \right) + o_p(1)\end{align*}
\normalsize

where $f^{\ast}(M\mid A, L)$ and ${b}^{\ast}_0(M,L)$ denote the limiting values of $\hat f(M\mid A, L)$ and $\hat{b}_0(M,L)$. This gives intuition to why the augmented inverse probability weighted estimator proposed in \citet{Fulcher2020} is consistent when models for $b_0(M,L)$ and $P(A=a\mid L)$ are correctly specified, or when the model for $P(M=m\mid A,L)$ is correctly specified.

Similarly, it can also be shown that as long as:
\begin{enumerate}
\item $\norm{\hat h_\dagger(L)-h_\dagger(L)}~ \norm{\hat f(a^\circ\mid L) - f(a^\circ\mid L)} = O_p(n^{-\nu})$, and
\item $\norm{\hat b_0(M,L)-b_0(M,L)}~ \norm{\hat f(a\mid L) - f(a\mid L)} = O_p(n^{-\nu}),~~\forall a$, and
\item $\norm{\hat b_0(M,L)-b_0(M,L)}~ \norm{\hat f(a\mid M, L) - f(a\mid M, L)} = O_p(n^{-\nu}),~~\forall a$
\end{enumerate}
for $\nu>1/2$ and where $\norm{f(x)} = \left\{\int |f(x)|^2dP(x)\right\}^{1/2}$, i.e. the $L_2(P)$ norm. Then, $\sqrt{n}\Big\{\Psi(\hat P) + P({\varphi}^{eff}(\hat{P})) - \Psi(P)\Big\} = o_p(1)$.
This can be accomplished, for example, if the nuisance functions are each consistently estimated at a rate of $n^{-1/4}$ or faster.

\bibliographystyle{apalike}
\bibliography{refs}